\documentclass[12pt]{iopart} 
\usepackage{latexsym}
\usepackage[T1]{fontenc}
\usepackage[utf8x]{inputenc}
\usepackage{latexsym}
\usepackage{times}
\usepackage{graphicx,amsfonts,amsbsy,amssymb,amsthm}
\usepackage{amsopn}
\usepackage{color}
\usepackage{cite}
\usepackage{hyperref}
\usepackage{braket}
\usepackage{float}
\usepackage{amssymb}
\usepackage{amscd}
\usepackage{amsfonts}
\usepackage{amstext}
\usepackage{amsthm}
\usepackage{algorithm}
\usepackage[noend]{algpseudocode}
\usepackage{url}
\usepackage{enumitem}

\usepackage{tikz}
\usetikzlibrary{fit,matrix}

\newcommand{\bx}{{\boldsymbol x}}
\newcommand{\by}{{\boldsymbol y}}

\newtheorem{theorem}{Theorem}
\newtheorem{lemma}[theorem]{Lemma}
\newtheorem{prop}[theorem]{Proposition}

\newtheorem{definition}[theorem]{Definition}
\newtheorem{remark}{Remark}

\def\={\;=\;} \def\+{\,+\,}

\def\mc {\mathcal}

\def\ms {\mathsf}

\def\Ketbra#1#2{\left|{#1}\vphantom{#2}\right\rangle\!\left\langle{#2}\vphantom{#1}\right|}

\begin {document}
\title[Can one hear a matrix?]{Can one hear a matrix? Recovering a real  symmetric matrix from its spectral data}
\author{Tomasz Maci\k{a}\.{z}ek}
\ead{tomasz.maciazek@bristol.ac.uk}
\address{School of Mathematics, University of Bristol, Fry Building, Woodland Road, Bristol BS8 1UG, UK}
\author{Uzy Smilansky}
\address{Department of Physics of Complex Systems, Weizmann Institute of Science, Rehovot 7610001, Israel}
\date{\today}

\begin {abstract} 
The spectrum of a real and symmetric $N\times N$ matrix determines the matrix up to unitary equivalence. More spectral data is needed together with some sign indicators to remove the unitary ambiguities. In the first part of this work we specify the spectral and sign information required for a unique reconstruction of general matrices. More specifically, the spectral information consists of the spectra of the $N$ nested main minors of the original matrix of the sizes $1,2,\dots,N$. However, due to the complicated nature of the required sign data, improvements are needed in order to make the reconstruction procedure feasible. With this in mind, the second part is restricted to banded matrices where the amount of spectral data exceeds the number of the unknown matrix entries. It is shown that one can take advantage of this redundancy to guarantee unique reconstruction of {\it generic} matrices, in other words, this subset of matrices is open, dense and of full measure in the set of real, symmetric and banded matrices.  It is shown that one can optimize the ratio between redundancy and genericity by using the freedom of choice of the spectral information input. We demonstrate our constructions in detail for pentadiagonal matrices.
\end{abstract}

\section {Introduction}
The answer to the question posed in the title is definitely: No!  The spectrum determines the matrix up to unitary equivalence. The question is then, what additional spectral  information  is required for this purpose. This question accompanies Mathematical Physics  already for a long time, and it is usually referred to as the spectral inversion or spectral reconstruction problems \cite{gladwell,marchenko}. The related reconstruction methods have a wide scope of applications in different areas of Physics, Materials Science and Engineering. They appear e.g., in studies of mechanical systems near equilibrium, where one tries to construct a quadratic Hamiltonian model \cite{Gantmacher 1960},  or in atomic, molecular and nuclear physics where spectra are measured, with the hope that they will provide information on the underlying interactions. In most cases the Hamiltonian operator is expressed as an ODE or PDE acting on an appropriate function spaces. A prime example is the Sturm Liouville operator which appears in various guises \cite{Gesztesy 2010}. In other cases, the systems are described in terms of discrete models, so that finite dimensional matrices are the operators under study. In the present paper we shall consider this case.  

Recent applications of inverse methods in experiments are numerous. We mention briefly a partial list. A model mechanical problem concerns a spring-mass system where the spring constants and masses can be deduced from the knowledge of system's eigenfrequencies \cite{gladwell,gladwell2}. More complex problems include studying vibrating beams \cite{gladwell84} or the composition of strings \cite{Cox 2012,PRT13}. Another paper \cite{DPZ20} studies neutrino oscillations by reconstructing the so-called mixing matrix in matter from the spectra of its main $(N-1)\times (N-1)$-minors. There, the relevant identity is the so-called eigenvector-eigenvalue identity \cite{tao}. Other types of eigenvector-eigenvalue identities have been used for parameter estimation in spin chain Hamiltonians \cite{Bur17} and Markov Models \cite{Bur18}.

In the present paper the spectral information about the matrix comes from the spectra of its $N$ nested main minors. Such a spectral information determines the matrix up to a finite ({\it a prori} large) number of possibilities. The reconstruction can be then made unique by providing additional discrete data encoded in the signs of certain expressions involving entries of the desired matrix. We devote much attention to the study of this sign problem. This may seem like a small subtlety, but the difficulty related to correctly choosing different types of signs in the study of the inverse matrix problems is often identified as one of the main factors that makes the practical use of the theoretical methods hard. G. M. L. Gladwell in his book \cite{gladwell} describes the subtle role played by different types of signs in the following way: ``{\it Signs, positive and negative, lie at the heart of any deep discussion of inverse problems.}''. 

Arguably, the reconstruction methods have been most successful for symmetric tridiagonal matrices often referred to as Jacobi matrices \cite{Gesztesy 2010}. For an $N$ dimensional Jacobi matrix, the number of unknown entries is $2N-1$ and the spectral information necessary for the reconstruction is the spectrum with $N$ eigenvalues, to which one adds the spectrum of the $N-1$ dimensional minor obtained by removing the last column and row from the original matrix (also called a main minor of order $N-1$). This data provides the diagonal elements (with their signs), but only the absolute value of the off diagonal entries can be  recovered  \cite{Hochstadt 1974, Hochstadt 1967, Hald1976}. Thus, the related sign problem is particularly simple -- it suffices to know the signs of the off-diagonal entries of the desired Jacobi matrix. 

When more general matrices which are less sparse are considered, the number of unknown entries is larger, and naturally, the number of input spectral data has to increase. This is accompanied  by the  sign problem becoming  more acute. The sign-indicators needed in the general case are not necessarily the signs of the individual entries. However, they are absolutely needed to resolve the ambiguity due to spectral invariance of the main minors under similarity with diagonal matrices with entries arbitrarily chosen from $\{-1,+1\}$. Various methods were developed for recovering banded or full matrices \cite{BG78,Konig81,BG87,CG-book,Loewy,Friedland,MW65,tao}. Most of them provide only an exemplary matrix satisfying the spectral data and stop short of resolving the sign-related or other ambiguities. One of the main approaches in reconstructing banded symmetric matrices is to fix the spectra of a few largest main minors (specifically $d+1$ main minors of orders $N$, $N-1$, $\dots$, $N-d$ for a $(2d+1)$-banded matrix) and using the Lanczos algorithm, bring the reconstructed matrix to a $(2d+1)$-banded form \cite{BG78,Konig81,BG87,CG-book}. Importantly, the Lanczos algorithm changes the spectra of all remaining main minors and thus this method differs from the approach presented in this work. Similar problems have been considered for matrices with complex entries. A reconstruction algorithm using spectra of all $2^{N}-1$ main minors has been developed \cite{GT06}. The algorithm produces an example of a complex matrix whose main minors have the given spectra and, under some regularity assumptions, the result is unique up to similarity transformations by nonsingular diagonal matrices and  the transposition operation \cite{Loewy}. Even more general settings concerning matrices with entries from an algebraically closed field have been considered. In particular, it has been shown that there exists a finite number of square $N\times N$ matrices with a prescribed spectrum and prescribed off-diagonal entries \cite{Friedland}. A different approach to the problem is presented in \cite{KMS16} where any Hermitian matrix can be reconstructed from its spectrum and the spectra of a suitable number of its perturbations. Recently, there has been some revived interest in the so-called eigenvector-eigenvalue identity \cite{tao} which allows one to find the amplitudes of eigenvector's entries of a hermitian matrix using the spectra of its $(N-1)\times (N-1)$-minors and the spectrum of the full matrix itself. As the authors of \cite{tao} point out, the eigenvector-eigenvalue identity appears in different guises in many places in the literature (see e.g. \cite{DNT86} -- the authors of \cite{tao} also provide a thorough review of the relevant literature). In the present paper we derive (\ref{appB}) an identity which bears some similarity to the eigenvector-eigenvalue identity. This identity was proved previously \cite{BG78,Konig81,BG87} using a different method.  

The present paper consists of three sections. The first  deals with the inverse problem of full, real and symmetric matrices of dimension $N$, where the number of unknown entries is $\frac{1}{2}N(N+1)$. The spectral data to be used is the union of the spectra of the first $N$ nested main minors of the matrix of the sizes $1,2,\dots,N$,  and $\frac{1}{2}N(N+1)$ sign indicators  needed for the complete reconstruction.  The  precise definition  of the sign indicators will be given below. The actual construction is inductive:  given a matrix $A$ and a minor $A^{(n)}$ of dimension $n$, its next neighbor $A^{(n+1)}$ is obtained by computing the $(n+1)$'th column from the given spectra  and sign indicators.  The uniqueness of the resulting solution is proved.  Thus, the matrix unfolds like a telescope, hence its name - the telescopic construction.  The fly in the ointment is that the computation of the sign indicators is rather cumbersome. 

The second section uses the telescopic method restricted to banded matrices with band width $D=2d+1$ much smaller than $N$.  The spectral input  exceeds the number of unknowns, but {\underline{this redundancy enables circumventing the need to compute the sign-indicators}}, and the only sign information required consists of the signs of the off-diagonal matrix elements in the diagonal which is $d$ -steps away from the main diagonal, just as it was the case for the Jacobi matrices. The proposed method is proved to provide a unique solution only for {\it generic} $D$-diagonal matrices. Namely, this subset is open, dense and of full measure in the set of real, symmetric and banded matrices specified by entries in ${\mathbb R}^{ ( N-\frac{1}{2})(d +1) } $.  An explicit criterion which distinguishes non-generic cases is proposed.  

 Finally, in the last section it is shown that the large redundancy which exists in the telescopic approach  can be reduced appreciably by studying a different construction:  One considers only the $d+1$ dimensional principal minors  $M^{(n)}$  which are distinguished by the position of their upper corner of $M^{(n)}$ along the diagonal. Their spectra and appropriate sign indicators are  used to compute  successive minors in  a recursive process.  This method will be referred to as the {\it sliding-minor construction}. Also in the sliding-minor construction we are able to introduce certain redundancy by considering $(d+2)$-dimensional sliding minors. Then, we show that the required sign data typically reduces to the signs of the off-diagonal matrix elements in the first row. The application of the last two methods for banded matrices, requires a special treatment of the upper $d\times d$ main minor, as will be explained in detail for each case. To allow a smooth flow of the main ideas, some of the more technical proofs of lemmas and theorems which are stated in the text are deferred to the appendices.

 We believe that turning the focus to generic rather than to unconditionally applicable methods increases the domain of practical application and the scope of the study of inverse problems.     

\subsection {Notations and preliminaries}
\label {notations}
Before turning to the main body of the paper, the present subsection  introduces  notations and conventions that will be used throughout. 
 
Let $A$ be a symmetric real $N\times N$ matrix. Denote by $A^{(n)}$ its $n\times n$ upper main diagonal minor, so that  $A^{(1)}= A_{1,1}$ and $A^{(N)}= A$. The upper suffix $(n)$ will be used throughout for the dimension of the subspace under discussion. 

The Dirac notation for vectors is used. That is, given a list $x^{(n)} = \{  x^{(n)}_j \}_{j=1}^n$, then
$\Ket{x^{(n)}}$  denotes the column vector in dimension $n$  with entries $ x^{(n)}_j$. Its transpose (row) will be denoted by
$\Bra{x^{(n)} }$ so that the scalar product is $\Braket{x^{(n)} | y^{(n)}}$. It is convenient also to introduce  the unit vectors $\Ket{e^{(n)}(j)},\ \ 1\le j\le n$ whose entries all vanish but for the $j$'th which is unity. Thus, clearly $\Braket{e^{(n)}(j) |x^{(n)}} = x^{(n)}_j  $, and $A^{(n )}_{i,j}:=\Braket{e^{(n )}(i) | A^{(n )} | e^{(n )}(j)}$. 
																									
\noindent For every minor $A^{(n)}, \ 1\le n\le(N-1)$ define the upper  off-diagonal part of the next  column    
\[\Ket{a^{(n)}} :=
\pmatrix{
A^{(n+1)}_{1,n+1} \cr 
\vdots \cr
A^{(n+1)}_{n,n+1}
} =\sum_{j=1}^n A^{(n+1)}_{j,n+1}  \Ket{e^{(n+1)}(j)} \ .
\]

\noindent The spectra of  $A^{(n)}$  for $1\le n \le N$ will be denoted by $\sigma^{(n)} =  \{\lambda^{(n)}_j \}_{j=1}^n $, with $\lambda^{(n)}_k \ge \lambda^{(n)}_j\  \forall \ k\ge j$, and the corresponding normalized eigenvectors (determined up to an overall sign) are  $\left\{\Ket{v^{(n)}(j) }\right\}_{j=1}^n$.  In general, the choice of eigenvectors is ambiguous. In the following, we will often  remove this ambiguity by fixing the choice of eigenvectors of $ A^{(n)}$. If the spectrum is non-degenerate, this is done, for instance by demanding the last entry of each $\Ket{v^{(n)}(j) }$ to be positive. For a fixed choice of eigenvectors, the projections of $  \Ket{a^{(n)}}$ on the eigenvectors of $  A^{(n )}$ will be denoted by $ s^{(n)}_j \xi^{(n)}_j$ where 
\begin{equation}
\hspace{-10mm}
\left\{\ s^{(n)}_j:= {\rm Sign} \left(\Braket{a^{(n)}|v^{(n)}(j) } \right)\ \right\}_{j=1}^n , \ {\rm  with \ \ \     Sign}(x)=\left \{ \begin{array}   {l}
  {+ 1\ {\rm for} \ x\ge 0} \\ {  -1\ {\rm for} \  x< 0 } \end{array}\right. \ .
\label{signind}
\end{equation}
  The  $ s^{(n)}_j$ will play the role of the sign-indicators in what follows. However, the arbitrary choice of the overall signs of the vectors $\Ket{v^{(n)}(j)}$ will not have any effect on the results,  as long as the chosen signs are not altered throughout the proof. 
 
% The only time we make use of the sign indicators is when we write a matrix in its spectral representation:
%\[A_{i, j}^{(N)} = \sum_{r=1}^N   \Braket{e^{(N)}(i)|v^{(N)}(r)}\lambda_r^{(N)} \Braket{v^{(N)}(r)|e^{(N)}(j)}.\]
%This depends exclusively on the relative signs of the entries of each eigenvector, and not on the arbitrary overall sign of the eigenvector. 

\section {Inversion by the telescopic construction}

Given the spectra $\sigma^{(n)}$ for $1\le n \le N$, one can use very simple arguments to derive the following information on the matrix $A$, without relaying on any sign information. They are known in different guises,   see e.g. references \cite{BG78,Konig81,BG87,CG-book}. We bring them here for completeness and as a background necessary for the ensuing developments. 

\begin{theorem}\label{thm1}
The spectra $\sigma^{(n)}$ for $1\le n \le N$ suffice to determine the diagonal elements 
of $A$ and the norms of the vectors  $|a^{(n)}\rangle$. 
\end{theorem}

\begin{proof} The minor $A^{(n+1)}$ is different from $A^{(n)}$ by the added diagonal entry $A^{(n+1)}_{n+1,n+1} $ to be denoted by $h$, and the off-diagonal column $\Ket{a^{(n)}}$ . Clearly   
\begin{equation}
\tr A^{(n+1)}-\tr A^{(n )}= A^{(n+1)}_{n+1,n+1} =h\ ,
\label {find h}
\end {equation} 
Hence $h$ is deduced directly from the spectral data. Also,
\begin{equation}\label{radius}
\fl \sum_{k=1}^n\left(A^{(n+1)}_{k,n+1}\right)^2 =\left|a^{(n)}\right|^2= \frac{1}{2}  \left( \tr\left( (A^{(n+1)})^2\right) - \tr\left( (A^{(n)})^2\right) -h^2  \right)=:R^{(n)}.
\end{equation} 
Thus  $h$,  and the value of $\left|a^{(n)}\right|^2$ can be computed for all $n$.
\end{proof}

\begin{remark}\label{rem1}
This  result can be considered as the inverse of the  Gerschgorin theorem \cite {gerschgorin}, which for symmetric matrices states that given a symmetric matrix $A$, its spectrum lies in the union of real intervals centered at the points $A_{n,n}$ and are of lengths $2\sum_{k=1} ^N |A_{n,k}|$. Theorem \ref{thm1} states that given the spectra of the successive minors, the diagonal elements of the matrix are determined precisely, and the vectors of off-diagonal elements  $a^{(a)}$ are restricted to a sphere of a radius determined by \Eref{radius}.    
\end{remark}
\begin{remark}\label{rem2}
	If $A$ is a Jacobi (symmetric, tridiagonal) matrix, only the $n$th component of $a^{(n)}$ is different from zero, and therefore the off-diagonal entries are determined up to a sign. In other words, one has to provide their signs so that the Jacobi matrix could be heard. 
\end{remark}

The following identities are similar in spirit to the former ones, and can be proven by similar ways. They will be used in the sequel.
    
\begin{eqnarray}\label{third}
\fl \Bra{a^{(n)}}A^{(n)}\Ket{a^{(n)}}= & \sum_{k=1}^n\lambda^{(n)}_k  |\langle a^{(n)}|v^{(n)}(k) \rangle|^2 = \sum_{k=1}^n\lambda^{(n)}_k | \xi^{(n)} _k| ^2 = \\ 
& =\nonumber \frac{1}{3}  \left( \tr[ (A^{(n+1)})^3] - \tr[ (A^{(n)})^3] -3h |a^{(n)}|^2 -h^3 \right )  \ .
\end{eqnarray}

and, 
\begin{equation}\label{determinants}
\hspace{-25mm}
\Bra{a^{(n)}}\left(A^{(n)}\right)^{-1}\Ket{a^{(n)}}=\sum_{k=1}^n\frac{1}{ \lambda^{(n)}_k}  \left|\Braket{a^{(n)}|v^{(n)}(k) }\right|^2 =\sum_{k=1}^n\frac{1}{ \lambda^{(n)}_k} |\xi^{(n)} _k| ^2   =  h-\frac{\det A^{(n+1)}}{\det A^{(n )}}\ ,
\end{equation}
which is valid if $0$ is not in the spectrum of $A^{(n)}$.
(If $0\in \sigma^{(n)} $ one can add to $A$ a constant multiple of the identity which renders all the spectra non-vanishing, without affecting  any of the proofs or  the results derived in the sequel.)

 The identities (\ref {radius},\ref{third},\ref{determinants}) are quadratic in the $n$ components of the vector $|a^{(n)}\rangle $. When $A$ is assumed to be a pentadiagonal ($D=5$), only the last two components of $|a^{(n)}\rangle $ do not vanish, and the quadratic forms describe three concentric curves - a circle and two conic sections in ${\mathbb R}^2$. Their mutual intersections are candidates for the solution for the inversion problem. This is discussed in detail in section (\ref {subsec:pentaconicsection}). 

\subsection{A general algorithm}\label{general}
Unlike the case of tridiagonal matrices where the signature indicators required are just the signs of the off diagonal entries (see Remark \ref {rem2}), for the  general case one requires    
the sign indicators  $\left\{\ s^{(n)}_j\right \}_{j=1}^n $  defined in (\ref{signind}) \ .

In the following, it will often be convenient to assume certain regularity properties of the matrix $A$. 
\begin{definition}[Regular matrix]\label{def:regular}
A real $N\times N$ symmetric matrix is called regular if and only if the following conditions are satisfied.
\begin{enumerate} 
\item The multiplicity of all eigenvalues of each $A^{(n)}$ is one. 
\item The successive spectra do not share a common value, $\sigma^{(n)}\cap\sigma^{(n+1)}=\emptyset$ for $n=1,\dots,N-1$.
\end{enumerate}
\end{definition}

\begin{remark}\label{per1}
The following theorem was kindly provided by Percy Deift \cite {percy1}, and it is quoted without proof.
\end{remark}

\begin{theorem} (Deift) The subset of regular matrices is open, dense and of full measure in the set of real and  symmetric  matrices. 	
 \end{theorem}     

\noindent Let $A$ be an $N\times N$ real symmetric and regular matrix. We prove 
\begin{theorem}\label{thm:main}
 Given the spectra  $\left \{\sigma^{(n)}\right \}_{n=1} ^N$  and  the sign indicators $\left \{s^{(n)}\right \}_{n=1} ^N$  as defined in (\ref {notations}) .  Then, these data suffice to construct the original matrix $A$ uniquely. 
 \end{theorem}
\noindent ( Irregular matrices will be discussed in a subsequent subsection ). 

\vspace{2mm}
 
\noindent Consider two successive minors $A^{(n)}$ and $A^{(n+1)}$. Cauchy's spectral interlacing theorem guarantees that
 \begin{equation}
 \lambda^{(n+1)}_1 \le \lambda^{(n)}_1 \le \lambda^{(n+1)}_2\ \le \cdots \le \   \lambda^{(n+1)}_n \le\lambda^{(n)}_n \le \lambda^{(n+1)}_{n+1}\ .
 \label{interlace}
 \end{equation}
\begin{lemma}\label{lemma:step}
 The vector $\Ket{a^{(n)}}$ and the diagonal element  $h:=A^{(n+1)}_{n+1,n+1}$\ , as well as  the (normalized) eigenvectors of the minor $A^{(n+1)}$, $\left\{ \Ket{v^{(n+1)}(j) } \right\}_{j=1}^{n+1}$,  are uniquely determined by the following data.
\begin{enumerate}
\item The full spectral data of $A^{(n)}$, i.e. its spectrum $\sigma ^{(n)}$ and eigenvectors $\left\{ \Ket{v^{(n)}(j) } \right\}_{j=1}^n$.
\item The spectrum $\sigma ^{(n+1)}$  of $A^{(n+1)}$,  
\item The sign indicators $s^{(n)}_j$, $j=1,\dots,n$.
\end{enumerate} 
\end{lemma}
\begin{proof}
The value of $h$ is determined by (\ref {find h}).  The next steps are based on the relation between the two spectra $\sigma ^{(n)}$ and $\sigma ^{(n+1)}$ which is used in the proof of  (\ref {interlace}). 
Define an auxiliary matrix $\tilde A^{(n+1)} $ with $ A^{(n)}$ being its main $n\times n$ minor,   $\tilde A^{(n+1)}_{n+1,n+1}=  h$, and whose off diagonal entries of the $(n+1)$'th column and row all vanish.  Then, the two matrices of dimension $(n+1)$,  $ A^{(n+1)}$ and $ \tilde A^{(n+1)}$ differ by a matrix of rank 2.
\begin{equation}
  A^{(n+1)} = \tilde A^{(n+1)} +\Ketbra{e^{(n+1)}(n+1)}{a^{(n)},0 }+\Ketbra{a^{(n)},0}{e^{(n+1)}(n+1)}
  \label{rank 2 pert} 
\end{equation} 
The spectrum  of $\tilde A^{(n+1)}$ consists of   $\sigma ^{(n)}\cup \{h\}$ . Correspondingly, the  eigenvectors of $\tilde A^{(n+1)}$ are obtained by increasing the dimension of each of the eigenvectors of $A^{(n)}$ by adding a $0$ in the $(n+1)$'th position and adding the eigenvector $\Ket{e^{(n+1)}(n+1)}$. These eigenvectors will be denoted by $\left\{\Ket{\tilde v^{(n+1)}(j)} \right\}_{j=1}^{n+1}$, where
\[ \hspace{-25mm} \Ket{\tilde v^{(n+1)}(j)}= \Ket{v^{(n)}(j),0}\quad{\rm for}\quad j=1,\dots,n\quad{\rm and}\quad     \Ket{\tilde v^{(n+1)}(n+1)}=\Ket{e^{(n+1)}(n+1)}.\] 

To express the spectrum and eigenvalues of $ A^{(n+1)}$, expand any of its eigenvectors in the basis $\left\{\Ket{\tilde v^{(n+1)}(j)} \right\}_{j=1}^{n+1}$, so that  
\begin{equation} 
\fl \Ket{v^{(n+1)}(k)}= \sum_{r=1}^{n+1} b_{k,r} \Ket{\tilde v^{(n+1)}(r)} \quad {\rm and }\quad A^{(n+1)}\Ket{v^{(n+1)}(k)}= \lambda^{(n+1)} _k \Ket{v^{(n+1)}(k)}.
\label{expand}   
\end{equation}
Replacing  $A^{(n+1)}$ by its form (\ref{rank 2 pert}) one easily finds that the expansion coefficients $ \{ b_{k,r} \}_{r=1}^{n+1}$  of the $k$th eigenvector satisfy
\begin{eqnarray}
\fl \lambda^{(n+1)}_k b_{k,r}&=& (1-\delta_{r,n+1}) b_{k,r}\lambda ^{(n)}_r +\delta_{r,n+1} h\  b_{k,n+1}\\ 
&+& \delta _{r,n+1}\sum_{j=1}^n b_{k,j} \Braket{a^{(n)},0 |\tilde v^{(n+1)}(j)} +(1-\delta_{r,n+1})\Braket{ \tilde v^{(n+1)}(r)|a^{(n)},0} b _{k,n+1}\ . \nonumber
\end{eqnarray}
From these relations and the definitions above we conclude that:
\begin{enumerate}
\item for $r=n+1$
\begin{equation}\label{rnp1}  
( \lambda^{(n+1)}_k -h ) b_{k,n+1} = \sum_{j=1}^n b_{k,j} \Braket{ a^{(n)} | v^{(n)}(j) }\ ,
\end{equation}
\item for $r\le n$
\begin{equation}  
( \lambda^{(n+1)}_k  -\lambda^{(n)}_r  ) b_{k,r} =  b_{k,n+1} \Braket{ v^{(n)}(r) |a^{(n)}}\ .
\label{rlessn} 
\end{equation}
\end{enumerate}
Since by the conditions of the theorem,   $ \lambda^{(n+1)}_k \ne \lambda^{(n)}_r $ for all the pairs $1\le r \le n,\  1\le k \le n+1 $, one can write
\begin{equation} 
b_{k,r} = \frac{\Braket{ v^{(n)}(r)|a^{(n)}}} {\lambda^{(n+1)}_k-\lambda^{(n)}_r}b_{k,n+1}  
\label{coeff}
\end{equation}
\begin{equation}
\lambda^{(n+1)}_k -h = \sum_{r=1}^n \frac{\left|\Braket{ a^{(n)}|v^{(n)}(r)}\right|^2} {\lambda^{(n+1)}_k-\lambda^{(n)}_r},  \ \ \ \forall\ \  1\le k \le n+1\ .
\label {spectrum}
\end{equation}
Moreover, all minors are assumed to be of full rank, and  both $\sigma^{(n+1)}$ and $\sigma^{(n)}$ are given. Therefore, one can consider 
 (\ref{spectrum}) as a set of linear equations from which the unknown $n$ parameters 
 \[(\xi^{(n)}_r)^2:=\left|\Braket{ a^{(n)}|v^{(n)}(r)}\right|^2,\quad r=1,\dots,n\]
 could be computed and none vanishes as will be shown in (\ref{xi-equation}) below. However, the number of equations exceeds by $1$ the number of unknowns, and therefore there exists a solution only if the equations are consistent. To prove consistency, we start by obtaining an explicit solution of the first $n$  equations in (\ref{spectrum}).  Introducing the Cauchy matrix
 \begin{equation}
 C_{k,r}=\frac{1}{\lambda^{(n+1)}_k-\lambda^{(n)}_r} , \ \ {\rm for} , \ \  1\le k,r \le n \ ,
 \end{equation}
the   unknown parameters $\{  (\xi^{(n)}_r) ^2 \}_{r=1}^n $ then read (see \ref{appB}): 
\begin{equation}\label{xi-equation}
(\xi^{(n)}_r)^2 = \sum_{k=1}^n \left (  C^{-1}\right )_{r,k}\     (\lambda^{(n+1)}_k -h)=-\frac{\prod_{k=1}^{n+1}\left(\lambda^{(n)}_r-\lambda^{(n+1)}_k\right)}{\prod_{k=1,k\neq r}^n\left(\lambda^{(n)}_r-\lambda^{(n)}_k\right)}. 
\end{equation}
 Formula (\ref{xi-equation}) can also be found in earlier works \cite{BG78,Konig81,BG87} and the recursive relation for the eigenvectors (\ref{coeff}) has been also used in \cite{Konig81}. Similarly, one can combine \Eref{coeff} with \Eref{xi-equation} to obtain (after requiring the eigenvectors of $A^{(n+1)}$ to be normalized to one) the following expression for the coefficients $ \left|b_{k,n+1}\right|^2$
 \begin{equation}\label{eigenvalue-eigenvector}
 \left|b_{k,n+1}\right|^2=\frac{\prod_{r=1}^{n}\left(\lambda^{(n+1)}_k-\lambda^{(n)}_r\right)}{\prod_{r=1,r\neq k}^n\left(\lambda^{(n+1)}_k-\lambda^{(n+1)}_r\right)}.
 \end{equation}
 \Eref{eigenvalue-eigenvector} is an instance of the eigenvector-eigenvalue identity from \cite{tao} which allows one to compute the amplitudes $\left|\Braket{e_{r}^{(n+1)}|\tilde v^{(n+1)}(k)}\right|$ directly from the spectra of the $n$ main minors of $A^{(n+1)}$ of the size $n\times n$. Here, we are using only one $n\times n$ minor (namely, $A^{(n)}$), thus we recover only the last squared entry of each eigenvector. By combining \Eref{eigenvalue-eigenvector} with \Eref{coeff} we can express the coefficients $b_{k,r}$ in terms of the spectra of $A^{(n)}$ and $A^{(n+1)}$, but the resulting formulae will be different from the eigenvector-eigenvalue identity   \cite{tao} as we use different spectral data and the entries refer to the basis of the eigenvectors of $A^{(n)}$ only.
 
The consistency of the linear $n+1$ equations (\ref{rnp1}) and (\ref{rlessn}) could be proved by substituting the $(\xi^{(n)}_r) ^2$ computed above in the  last equation in (\ref{spectrum}), and showing that the resulting equation  
\begin{equation}
 (\lambda^{(n+1)}_{n+1} -h ) = \sum_{r=1}^n\sum_{k=1}^n  \frac{1}{\lambda^{(n+1)}_{n+1}-\lambda^{(n)}_r} \   \left (  C^{-1}\right )_{r,k}  (\lambda^{(n+1)}_k -h) , 
 \label{consistent}
\end{equation}
 is satisfied identically.
An explicit expression for $  \left (  C^{-1}\right )_{r,k}$ is given in   \cite{Gow 1992} : 
\begin{eqnarray}
  \left (  C^{-1}\right )_{r,k}&=&\frac{p( \lambda^{(n)}_r ) q(\lambda^{(n+1)}_k)}
  {(\lambda^{(n+1)}_r-\lambda^{(n)}_k) p\ '(\lambda^{(n+1)}_k)q\ '(\lambda^{(n)}_r)},\nonumber \\   
   {\rm where : }\ p(x)&=&\prod_{i=1}^n(x-\lambda^{(n+1)}_i)\   ;\   
   q(x)=\prod_{i=1}^n(x-\lambda^{(n)}_i) \ .
\end{eqnarray}
The identities which are proved in \ref{appA} show that  equation (\ref{consistent}) is satisfied identically provided that  $h=\sum_{k=1}^{n+1}\lambda^{(n+1)}_k -\sum_{r=1}^{n}\lambda^{(n )}_r $ as is the case in the present context. 
  
 To recapitulate, the solution of equations (\ref{spectrum}) exists and it provides the $(\xi^{(n)}_r)^2$ in terms of the spectra in a unique way as shown in \Eref{xi-equation}. This information together with the sign indicators  determine  $ \langle a^{(n)}|v^{(n)}(r)\rangle = s^{(n)}_r\xi^{(n)}_r  $, and the vector $\Ket{a^{(n)}}$ is obtained by an orthogonal change of basis. Moreover, the possible $2^n$ combinations of the sign indicators encapsulate all possible choices of the overall signs of the eigenvectors of $A^{(n)}$. Thus, an {\it a priori} choice of the overall phases of the eigenvectors does not result with any loss of generality or the loss of solutions.
  \end{proof}  

\begin{remark}\label{carlos}
A complementary result to the above was recently reported in  \cite {carlos1}. Using another approach it is proved that given a real symmetric matrix $\tilde A^{(n)}$ whose spectrum $\tilde \sigma ^{(n)}$ is simple (no multiplicities) and a list $\mu^{(n+1)}$  of $(n+1)$ arbitrary numbers which interlace with  $\tilde\sigma ^{(n)}$, one can construct a column vector $|\tilde a^{(n)}\rangle$ and a real $\tilde h$  which, when used to complete $\tilde A{(n)}$ to  $\tilde A^{(n+1)}$ yields $\tilde A^{(n+1)}$ with the spectrum $\mu^{(n+1)}$. Moreover, $|\tilde a^{(n)}\rangle$ can be constructed uniquely once an orthant (whose choice provides the sign indicators) is prescribed in the $n$ dimensional space spanned by the eigenvectors of $\tilde A^{(n)}$. It is also proved that the assignment of $|\tilde a^{(n)}\rangle$ and $\tilde h$ to $\mu^{(n+1)}$ is a homeomorphism onto its image and a differomorphism.
\end{remark} 
 
 The knowledge  of the  $\Braket{a|v^{(n)}(r)}$ with their signs enables the use of (\ref{coeff}) to obtain the coefficients of the eigenvectors of $A^{(n+1)}$ up to the constant factor $b_{k,n+1}$ which can be determined by requiring a proper normalization.

\begin{proof} (of Theorem \ref{thm:main})
The proof proceeds by a recursive construction: Start with a known $A^{(1)}$
and with the known spectrum of  $A^{(2)}$. Clearly, $\lambda^{(1)}_1=A_{1,1},\ {\rm and} \ v^{(1)}=1$. Furthermore,

%The spectrum of $A^{(2)}$ is
%\begin{equation}
%\lambda^{(2)}_{1,2}=\frac{1}{2} \left ( A_{1,1}+A_{2,2} \pm \sqrt{(A_{1,1}-A_{2,2})^2 +4 A_{1,2}^2 }\right ) .
%\end{equation}
 \[\fl h=\lambda^{(2)}_1+\lambda^{(2)}_2 - \lambda^{(1)}_1 =A_{2,2},\quad a=A_{1,2},\quad \Braket{a| v^{(1)}}= A_{1,2},\quad s^{(1)}_1 = \frac{A_{1,2}}{|A_{1,2}|}. \]
 Hence, \Eref{xi-equation} gives
 \[\fl (\xi^{(1)}_{1}) ^2=\left(A_{1,2}\right)^2=\left(\lambda_1^{(1)}-\lambda^{(2)}_1\right)\left(\lambda^{(2)}_2-\lambda_1^{(1)}\right)=\left(A_{1,1}-\lambda^{(2)}_1\right)\left(\lambda^{(2)}_2-A_{1,1}\right).\]
 Therefore, we get two possible solutions for $A^{(2)}$ corresponding to choosing $s^{(1)}_1=+1$ or $s^{(1)}_1=-1$, which written explicitly read
 \begin{equation}\label{a2}
 \fl A^{(2)}=
 \pmatrix{
A_{1,1} & s^{(1)}_1\sqrt{\left(A_{1,1}-\lambda^{(2)}_1\right)\left(\lambda^{(2)}_2-A_{1,1}\right)} \cr
s^{(1)}_1\sqrt{\left(A_{1,1}-\lambda^{(2)}_1\right)\left(\lambda^{(2)}_2-A_{1,1}\right)} & \lambda^{(2)}_1+\lambda^{(2)}_2 - A_{1,1} }.
 \end{equation}
The normalized eigenvectors can be computed from \Eref{coeff}. In the standard basis, their forms read
 \begin{equation}\label{a2-eigenv}
 \fl
 \Ket{v^{(2)}(1)}=\frac{1}{\sqrt{1+\alpha^2}}
 \pmatrix{
 \alpha \cr
 1
 }
 ,\
 \Ket{v^{(2)}(2)}=\frac{1}{\sqrt{1+\alpha^2}}
 \pmatrix{
-1 \cr
  \alpha
 }
 ,\ 
 \alpha:=-s^{(1)}_1\sqrt{\frac{\lambda^{(2)}_2-A_{1,1}}{A_{1,1}-\lambda^{(2)}_1}}.
 \end{equation}

 Thus the theorem is valid for $n=1,2$. Note also that one can show that (\ref{spectrum}) is satisfied - which is not clear at first sight but can be easily checked. The lemma provides the final stage of the inductive proof.
\end{proof}

The above inductive procedure is optimal for generic matrices whose all entries are typically nonzero. However, the requirement of providing signs of projections of the unknown column to the eigenvectors of $A^{(n)}$ seems awkward for practical use. However, when one wants to apply this result to banded matrices, it turns out that there is large  redundancy which can be used to our advantage. In particular, as we argue in the following sections, the inductive reconstruction procedure for banded matrices can be improved so that the number of required sign data can be tremendously reduced.

\subsubsection{Explicit reconstruction of $A^{(3)}$.}
\label{3x3-example}
It is  instructive to work out  the above inductive procedure explicitly for $n=3$. The results  will prove useful in the following sections.

\noindent Consider the problem of reconstructing a regular $3\times 3$ real symmetric matrix, $A^{(3)}$, given its spectrum $\lambda_1^{(3)}<\lambda_2^{(3)}<\lambda_3^{(3)}$ and its $2\times 2$ top-left minor  $A^{(2)}$. It has a spectral decomposition
\[A^{(2)}=\lambda_1^{(2)}\Ketbra{v^{(2)}(1)}{v^{(2)}(1)}+\lambda_2^{(2)}\Ketbra{v^{(2)}(2)}{v^{(2)}(2)},\quad \lambda_1^{(2)}<\lambda_2^{(2)},\]
where
\[
\lambda^{(2)}_{1,2}=\frac{1}{2} \left ( A^{(2)}_{1,1}+A^{(2)}_{2,2} \pm \sqrt{(A^{(2)}_{1,1}-A^{(2)}_{2,2})^2 +4 \left(A^{(2)}_{1,2}\right)^2 }\right ) .
\]
and
 \begin{equation}\label{m2-eigenv}
 \fl
 \Ket{v^{(2)}(1)}=\frac{1}{\sqrt{1+\alpha^2}}
 \pmatrix{
 \alpha \cr
 1
 }
 ,\quad
 \Ket{v^{(2)}(2)}=\frac{1}{\sqrt{1+\alpha^2}}
 \pmatrix{
-1 \cr
  \alpha
 }
 ,\quad
 \alpha:=-s\sqrt{\frac{\lambda^{(2)}_2-A^{(2)}_{1,1}}{A^{(2)}_{1,1}-\lambda^{(2)}_1}}
 \end{equation}
 with $s={\mathrm{Sign}}\left(A^{(2)}_{1,2}\right)$.
The unknown entries of matrix $A^{(3)}$ read
\begin{equation}\label{m3a}
\hspace{-20mm}
A^{(3)}_{i,3}=s^{(2)}_1|\xi_1^{(2)}|\left(v ^{(2)}(1)\right)_i+s^{(2)}_2|\xi_2^{(2)}|\left(v^{(2)}(2)\right)_i,\quad s^{(2)}_i\in\{+1,-1\},\quad  i\in\{1,2\},
\end{equation}
where $\{\xi^{(2)}_r\}_{r=1}^2$ are given by \Eref{xi-equation}. Thus, there are four possible solutions for the last column of $A^{(3)}$ corresponding to different choices of signs $s_i$. Next, let us show how to determine $s_i$ directly from the sign of a certain expression involving only matrix elements of  $A^{(3)}$. Using  \Eref{xi-equation} and \Eref{m2-eigenv}, we get that
\begin{eqnarray*}
\sqrt{1+\alpha^2}A^{(3)}_{1,3}=s_1|\xi_1^{(2)}|\alpha-s^{(2)}_2|\xi_2^{(2)}|,\quad \sqrt{1+\alpha^2}A^{(3)}_{2,3}=s^{(2)}_1|\xi_1^{(2)}|+\alpha s_2|\xi_2^{(2)}|.
\end{eqnarray*}
This allows us to express $s^{(2)}_1$ and $s^{(2)}_2$ as
\begin{eqnarray*}
s^{(2)}_1=\frac{\alpha A^{(3)}_{1,3}+A^{(3)}_{2,3}}{\sqrt{1+\alpha^2}|\xi_1^{(2)}|},\quad s^{(2)}_2=\frac{\alpha A^{(3)}_{2,3}-A^{(3)}_{1,3}}{\sqrt{1+\alpha^2}|\xi_2^{(2)}|}.
\end{eqnarray*}
Because the denominators of the above expressions are positive numbers, we finally get
\begin{eqnarray} \label{signs2to3}
s^{(2)}_1={\mathrm{Sign}}\left(-sA^{(3)}_{1,3}|\alpha|+A^{(3)}_{2,3}\right), s^{(2)}_2=-{\mathrm{Sign}}\left(-sA^{(3)}_{2,3}|\alpha|+A^{(3)}_{1,3}\right).
\end{eqnarray}
The main advantage of these expressions is that they provide  expressions for the signatures $s^{(2)}_1,s^{(2)}_2$ in terms of data directly available from the matrix.    
%\end{example}%

\subsection{Non-regular matrices}

\begin{figure}[h]
\centering
\includegraphics[width=.7\textwidth]{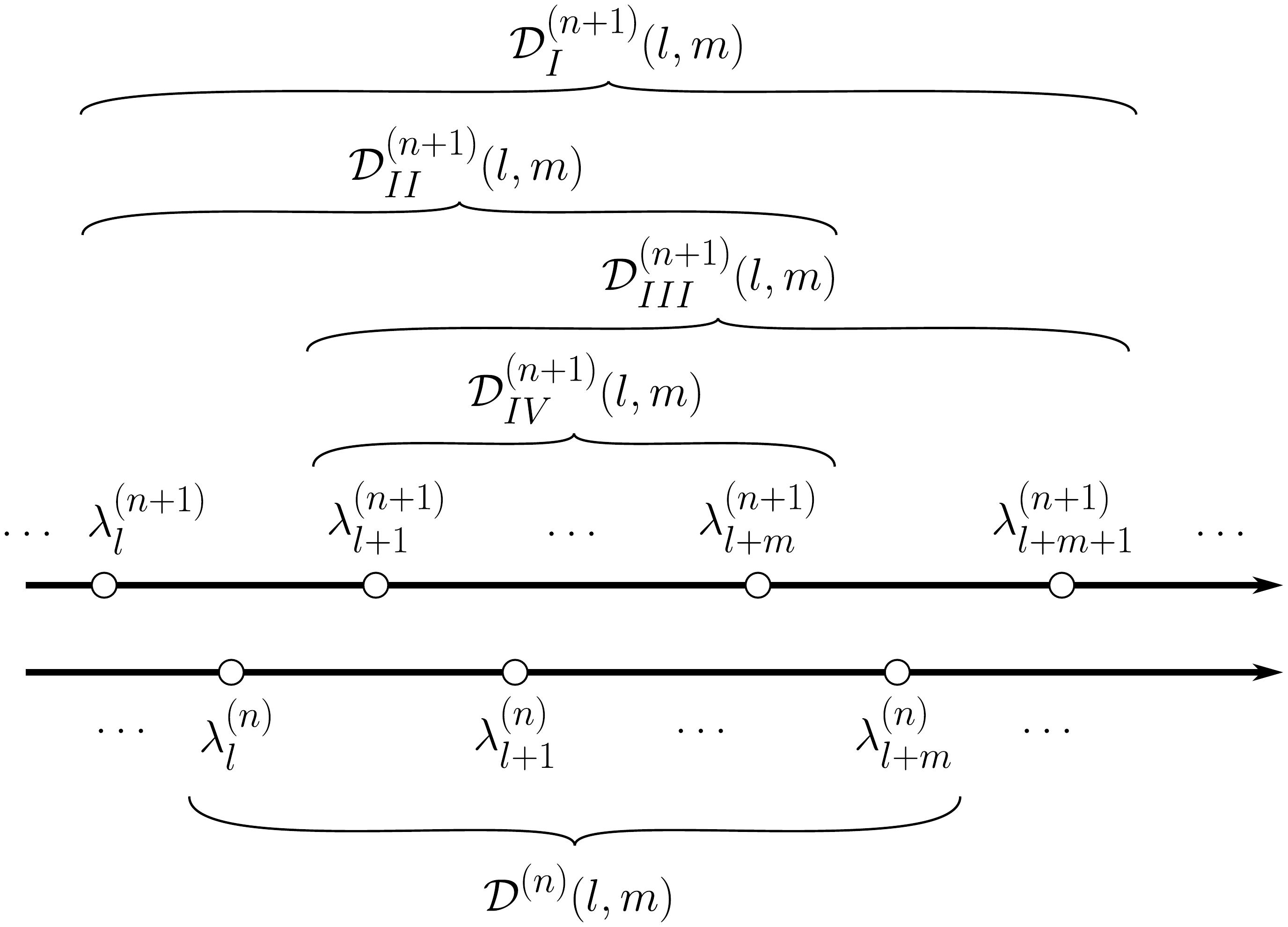}
\caption{The four possibilities for the position of the degeneracy indices $D^{(n+1)}(l,m)$ relative to $\mc{D}^{(n)}(l,m)$.} 
\label{degeneracy}
\end{figure}

In this subsection we show that non-regular matrices that have degeneracy in $\sigma^{(n)}$ or where the overlaps $\sigma^{(n)}\cap \sigma^{(n+1)}$ are nonempty can be effectively treated using the methods developed above for regular matrices. We consider the situation where there is a single block of degeneracy in $\sigma^{(n)}$, i.e. for some $1\leq l\leq n$ and $m\geq 0$ we have $\lambda_l^{(n)}=\lambda_{l+1}^{(n)}=\dots=\lambda_{l+m}^{(n)}:=\lambda$ and other eigenvalues of $A^{(n)}$ are non-degenerate. Let us denote the respective degeneracy indices by
\[\mc{D}^{(n)}(l,m):=\{l,l+1,\dots,l+m\}.\]
The interlacing property (\ref{interlace}) dictates the following possibilities for the respective degeneracy indices $D^{(n+1)}(l,m)$ (see Fig.~\ref{degeneracy}):
\begin{eqnarray*}
\fl \mc{D}_{I}^{(n+1)}(l,m):=\{l,l+1,\dots,l+m+1\},\\ 
\fl \mc{D}_{II}^{(n+1)}(l,m):=\{l,l+1,\dots,l+m\},\\
\fl \mc{D}_{III}^{(n+1)}(l,m):=\{l+1,l+2,\dots,l+m+1\}, \\
\fl \mc{D}_{IV}^{(n+1)}(l,m):=\{l+1,l+2,\dots,l+m\}\quad {\mathrm{for}} \quad m\geq 1,\quad \mc{D}_{IV}^{(n+1)}(l,0):=\emptyset.
\end{eqnarray*}

%These relations go  both ways, that is, any one of the  cases $I-IV$, or  the $m=0$ case,  for the spectrum $\sigma ^{n+1}$ imply the degeneracy $D^{(n)}(l,m)$ in $\sigma^{(n)}$. 
In what follows we assume that the eigenvalues $\lambda_{k}^{(n+1)}$ with $k\notin \mc{D}^{(n+1)}(l,m)$ are non-degenerate. Furthermore, the degenerate eigenspaces of $A^{(n)}$ and $A^{(n+1)}$ will be denoted by $V^{(n)}(\lambda)$ and $V^{(n+1)}(\lambda)$ respectively, and their respective orthogonal complements in ${\mathbb R}^{n}$ and ${\mathbb R}^{n+1}$ by $V^{(n)}(\lambda)^\perp$ and $V^{(n+1)}(\lambda)^\perp$. We will also embed ${\mathbb R}^{n}$ into ${\mathbb R}^{n+1}$ by appending a zero at the end of every vector. With a slight abuse of notation we will also denote by $V^{(n)}(\lambda)$ and $V^{(n)}(\lambda)^\perp$ the images of these spaces under the above embedding. 

Theorem \ref{thm:degenerate} below shows how to reduce each of the above four degenerate cases to a regular problem, where Theorem \ref{thm:main} can be applied. Interestingly, each case requires a different procedure.

\begin{theorem}\label{thm:degenerate}
Assume that the spectra of $A^{(n)}$ and $A^{(n+1)}$ as well as the eigenvectors $\left\{\Ket{v^{(n)}(r)}\right\}_{r=1}^n$ of $A^{(n)}$ are given and fixed. Assume that the spectra have single degeneracy blocks with the eigenvalue $\lambda$ given by degeneracy indices $\mc{D}^{(n)}(l,m)$ and $\mc{D}^{(n+1)}(l,m)=\mc{D}_i^{(n+1)}(l,m)$ for some $i\in\{I,II,III,IV\}$ respectively. The matrix $A^{(n+1)}$ is reconstructed from the above spectral data as follows.
\begin{enumerate}[label=\Roman*.]
\item The vector $\Ket{a^{(n)}}$ as well as the non-degenerate eigenvectors of $A^{(n+1)}$ belong to the space $V^{(n)}(\lambda)^\perp$. They are constructed by applying Theorem \ref{thm:main} to the truncated regular matrices $\tilde A^{(n-m)}:=A^{(n+1)}\bigg{|}_{V^{(n)}(\lambda)^\perp}$ and $\tilde A^{(n-m-1)}:=A^{(n)}\bigg{|}_{V^{(n)}(\lambda)^\perp}$. The degenerate eigenvectors of $A^{(n+1)}$
\[\Ket{\tilde v^{(n+1)}(k)}=b_{n+1,k}\Ket{e^{(n+1)}}+\sum_{r=1}^nb_{k,r} \Ket{v^{(n)}(r),0},\quad k\in \mc{D}_{I}^{(n+1)}(l,m)\]
have coefficients
\[b_{k,r}=\frac{\Braket{ a^{(n)}|v^{(n)}(r)}}{\lambda-\lambda_r^{(n)}}b_{k,n+1},\quad r\notin \mc{D}^{(n)}(l,m), \quad k\in \mc{D}_{I}^{(n+1)}(l,m).\] 
The remaining coefficients $b_{k,n+1}$ and $b_{r,k}$ with $k\in \mc{D}_I^{(n+1)}(l,m)$ and $r\notin \mc{D}^{(n)}(l,m)$ can be arbitrary so that the resulting eigenvectors form an orthonormal set.
\item The vector $\Ket{a^{(n)}}$ as well as the non-degenerate eigenvectors of $A^{(n+1)}$ belong to the space $V^{(n)}(\lambda)^\perp$. They are constructed by applying Theorem \ref{thm:main} to the truncated regular matrices $\tilde A^{(n-m)}:=A^{(n+1)}\bigg{|}_{V^{(n)}(\lambda)^\perp}$ and $\tilde A^{(n-m-1)}:=A^{(n)}\bigg{|}_{V^{(n)}(\lambda)^\perp}$. The degenerate eigenvectors of $A^{(n+1)}$ form an orthonormal basis of $V^{(n)}(\lambda)$.
\item The same as case II above.
\item The space $V^{(n+1)}(\lambda)$ is a subspace of $V^{(n)}(\lambda)$ of codimension one. The degenerate eigenvectors of $A^{(n+1)}$, $\Ket{\tilde v^{(n+1)}(k)}$, $k\in \mc{D}^{(n+1)}(l,m)$ can be freely chosen as any orthonormal subset of $V^{(n)}(\lambda)$. They determine the embedding $\phi:\ V^{(n+1)}(\lambda)\subset V^{(n)}(\lambda)$. The vector $\Ket{a^{(n)},0}$ as well as the remaining eigenvectors of $A^{(n+1)}$ belong to the orthogonal complement of the image of the embedding $\phi$ in ${\mathbb{R}}^{n+1}$, $\phi(V^{(n+1)}(\lambda))^\perp$. They are constructed using Theorem \ref{thm:main} applied to the regular matrices $\tilde A^{(n-m+1)}:=A^{(n+1)}\bigg{|}_{V^{(n+1)}(\lambda)^\perp}$ and $\tilde A^{(n-m)}:=A^{(n)}\bigg{|}_{\phi(V^{(n+1)}(\lambda))^\perp}$.
\end{enumerate}

\end{theorem}
\noindent The proof of Theorem \ref{thm:degenerate} is deferred to \ref{appC}.

%In the case when there are multiple blocks of degeneracy in $A^{(n)}$ and $A^{(n+1)}$, the relevant regular truncated matrices are constructed according to the rules stated in Theorem \ref{thm:degenerate} applied to each degeneracy block.

\section{Applications to banded matrices}

In the present chapter we shall discuss the application of the above formalism for banded matrices. The fact that the spectral data exceeds the number of unknowns will be  shown to provide  stringent constraints on the sign distribution of the off diagonal elements, so that generically their signs are determined up to an overall sign per column. If not stated otherwise, we will assume that the considered matrices are regular.

\subsection {Tridiagonal matrices} 
 In Remark \ref{rem1} above we have shown that  the present construction applies for the tridiagonal case. However it is certainly not efficient since the two spectra and the signs of the off-diagonal matrix elements suffice for the purpose - as is well known \cite{Anderson 1970}. 

\subsection{Pentadiagonal matrices}\label{subsec:penta-main-minors}
The next simple case which illustrate however the main ingredients of the general case are the $D=5$ - banded matrices a.k.a. pentadioagonal matrices.  
Here, 
 $$a^{(n)}_1=a^{(n)}_{2}= \cdots = a^{(n)}_{n-2} =0\ ,$$ 
  hence the inductive step requires solving a number of equations in the two real variables $a^{(n)}_{n }\  {\rm and}\ a^{(n)}_{n-1}.$
We shall address the subject using two different approaches. In the first we shall apply the method based on Theorem \ref{thm:main}. In the second  we shall use the quadratic forms (\ref{third},\ref{determinants}), which give an alternative point of view that applies exclusively for pentadiagonal matrices. 

Note that for both approaches, one should compute the upper  $3 \times 3 $ minor using the procedure  outlined in subsection (\ref{3x3-example}). Alternatively, it suffices to provide as input the spectrum and the eigenvectors of the upper $3 \times 3 $ minor, and use this data as the starting data for the ensuing inductive steps.  
     
\subsubsection {Applying Theorem \ref{thm:main} to the pentadiagonal case}
 In this subsection, we will focus on answering the following two questions
\begin{enumerate}
	\item Given that the spectra $\sigma^{(n)}$, $n=1,\dots,N$ correspond to a pentadiagonal matrix, what additional information is needed to uniquely reconstruct the matrix?
	\item What are the necessary conditions for a matrix $A^{(n)}$ and the spectrum $\sigma^{(n+1)}$ of $A^{(n+1)}$ so that $A^{(n+1)}$ is a  pentadiagonal matrix?
\end{enumerate}

Starting with  the first question, the general answer is given by Theorem \ref{thm:main}. However, we will make use of the redundant information to show that typically, one can do away with the computation of the sign indicators as prescribed in the theorem. It will be shown that the vectors $|a^{(n)} \rangle$ are determined up to an overall sign, which is uniquely provided by the easily available signs of the entries in the upper diagonal.

Consider the inductive step, where $A^{(n)}$ and the spectrum $\sigma^{(n+1)}$ are known. All vectors $\ket{a^{(n)}}$ have the same norm determined by \Eref{radius}. Hence,
\[\fl \left(a^{(n)}_{n-1}\right)^2+\left(a^{(n)}_{n}\right)^2=\left(R^{(n)}\right)^2:=\frac{1}{2}  \left ( \tr[ (A^{(n+1)})^2] - \tr[ (A^{(n)})^2] \right ) -h^2  ).\]
Furthermore, the solutions of \Eref{xi-equation} imply that
\begin{equation}\label{lines}
\fl \textrm{either}\  a^{(n)}_{n-1}v^{(n)}(r)_{n-1}+a^{(n)}_{n}v^{(n)}(r)_{n}= \xi_r^{(n)} \ \textrm{or}\quad a^{(n)}_{n-1}v^{(n)}(r)_{n-1}+a^{(n)}_{n}v^{(n)}(r)_{n}=- \xi_r^{(n)} 
\end{equation}
for every $r=1,\dots,n$. Thus, every $\ket{a^{(n)}}$ belongs to the intersection of one of the above lines (for every $r$) with the circle of radius $R^{(n)}$ (recall also that $\xi_r^{(n)}$ were defined to be  positive).

\begin{figure}[h]
\centering
\includegraphics[width=.45\textwidth]{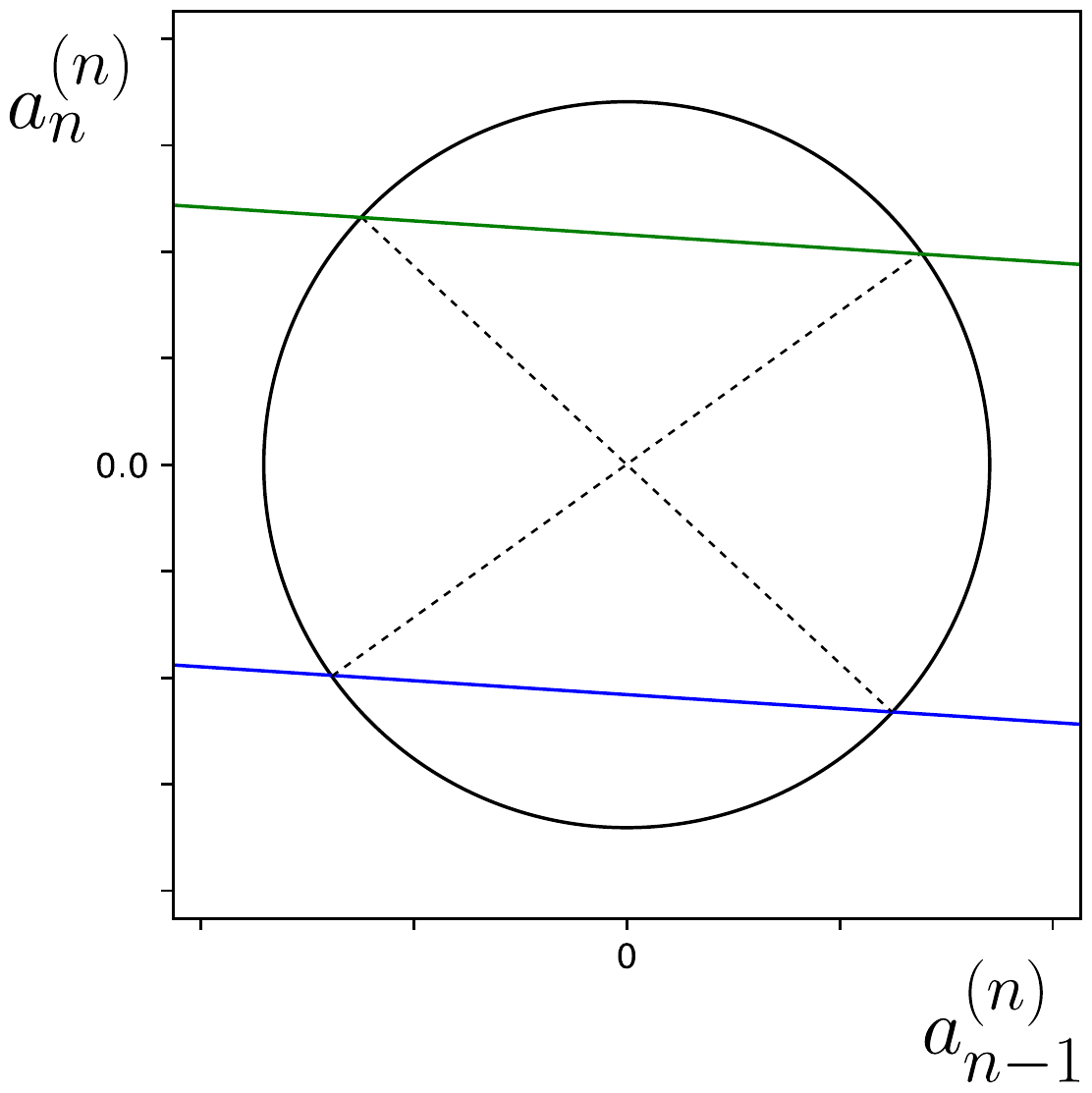}
\caption{Intersecting lines from \Eref{lines} with the circle. The blue line corresponds to the line with $ \xi_r^{(n)} $ as the constant while the green line corresponds to the choice of $- \xi_r^{(n)}$. Dashed lines connect the antipodal solutions.} 
\label{circle-intersection}
\end{figure}

By the assumption of Theorem \ref{thm:main},  $\sigma^{(n)}$ is non-degenerate and  none of the elements of $\sigma^{(n)}$ belongs to $\sigma^{(n+1)}$. By \Eref{xi-equation}, this implies that $\xi_r ^{(n)} > 0$ for all $r=1,\dots,n$. Then, \Eref{lines} defines two distinct parallel lines for each $r$. Moreover, for a fixed $r$, the intersection of the two lines with the circle occurs at two distinct pairs of points, where a single pair consists of a point and its antipode (see Fig.~\ref{circle-intersection}).

\begin{figure}[h]
\centering
\includegraphics[width=.45\textwidth]{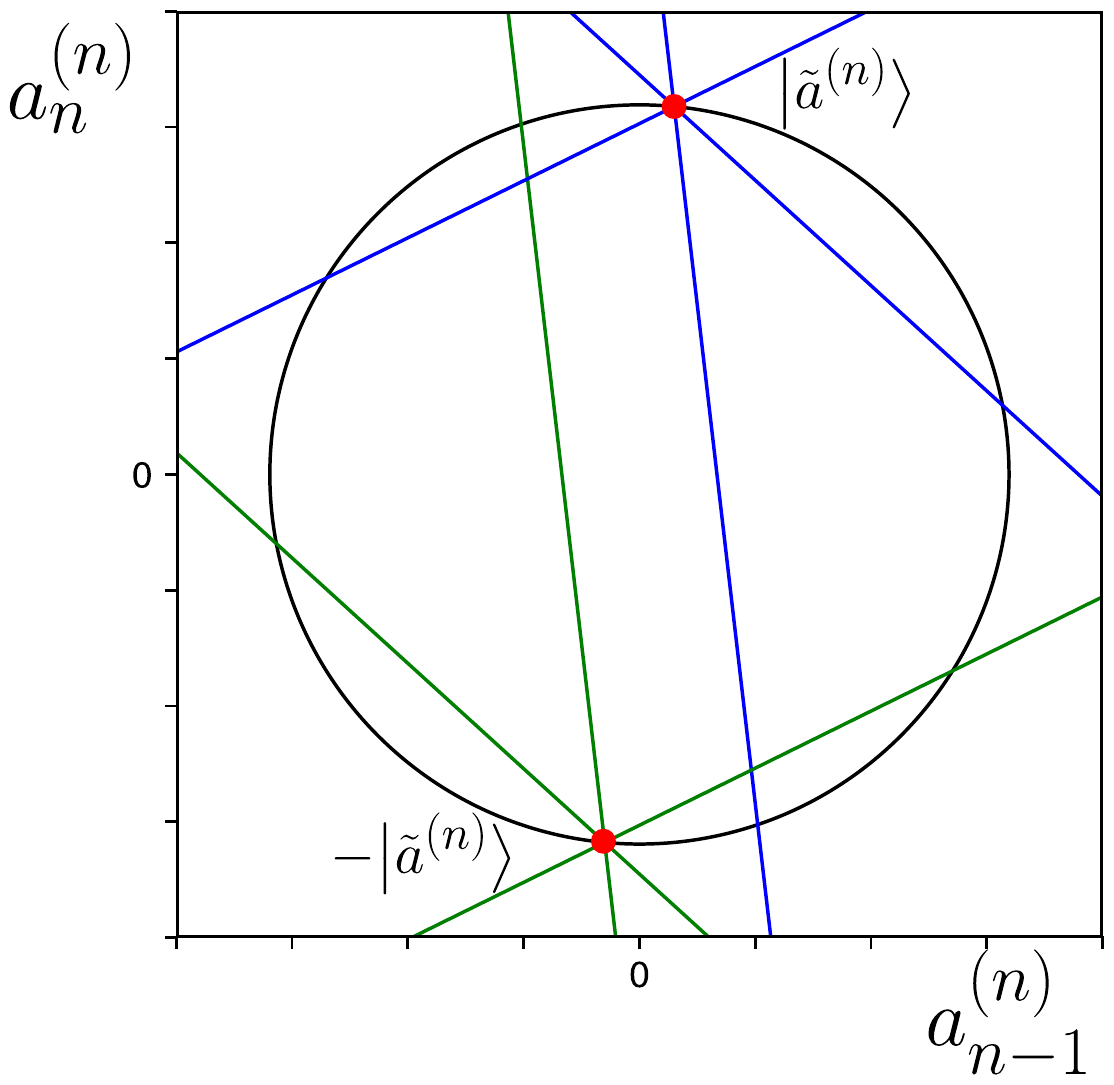}
\caption{A typical situation showing the result of intersecting lines from \Eref{lines} with the circle for $r=1,2,3$. Blue lines correspond to the line with $ \xi_r^{(n)} $ as the constant while green lines correspond to the choice of $- \xi_r^{(n)} $.}
\label{circle-intersection-generic}
\end{figure}

Since  the spectral data comes from a  pentadiagonal matrix, each of the lines from \Eref{lines} must have an intersection  point $\ket{\tilde a^{(n)}}$ or $-\ket{\tilde a^{(n)}}$ where $\ket{\tilde a^{(n)}}$ is the $n$th column  (with only two non zero entries) of the original matrix (see Fig.~(\ref{circle-intersection-generic}).

\begin{figure}[h]
\centering
\includegraphics[width=.45\textwidth]{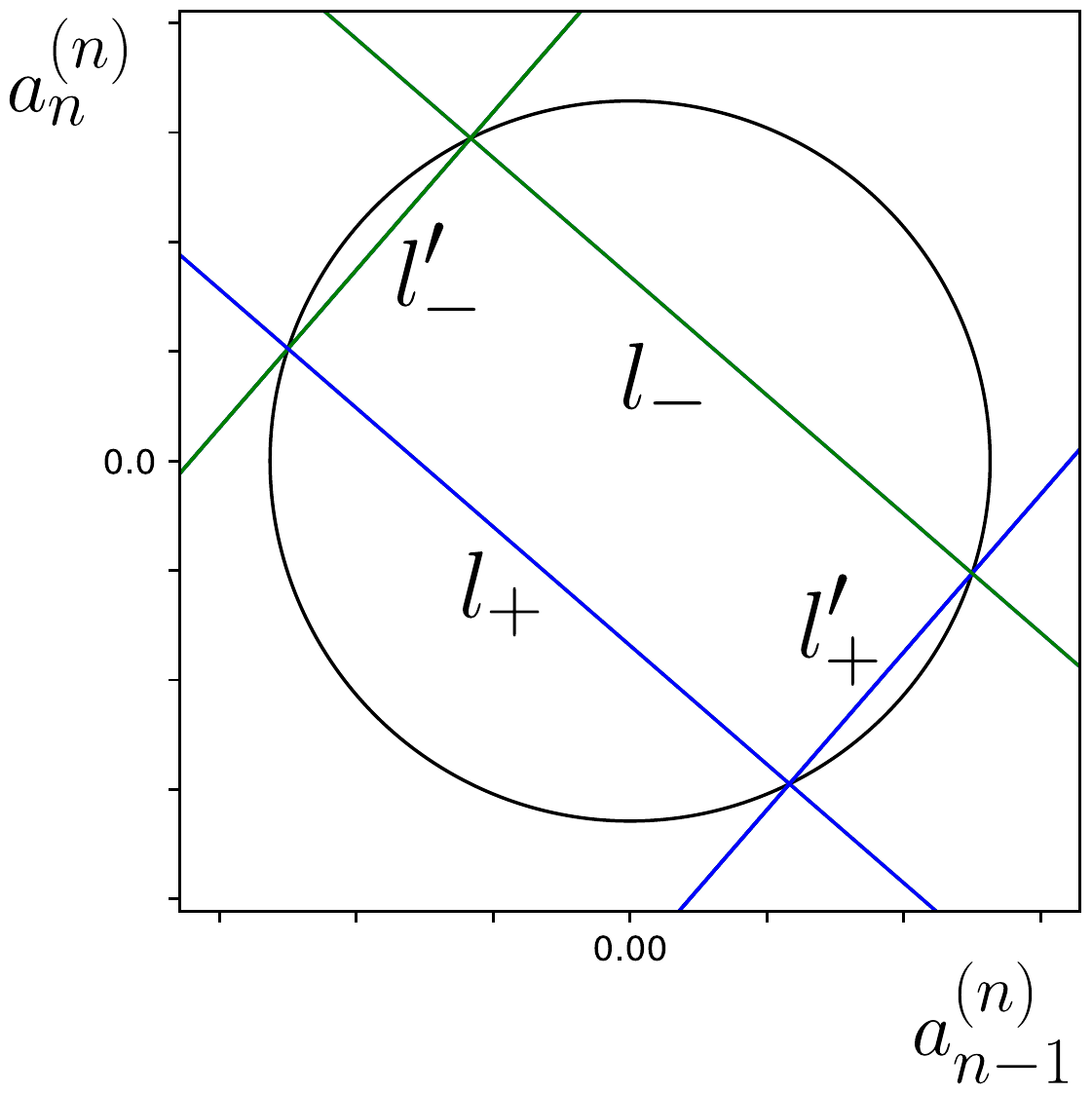}
\caption{A typical situation where intersecting all lines from \Eref{lines} with the circle leads to four solutions for $\ket{a^{(n)}}$.}
\label{intersections-degenerate}
\end{figure}

However, there exist special situations when such intersections with the circle give more than just one solution up to a sign. These are highly degenerate cases where all lines are determined by a fixed pair of parallel lines $(l_+,l_-)$. More precisely, for every $r$, the lines from \Eref{lines} are either equal to $(l_+,l_-)$ or to another fixed pair of lines $(l'_+,l'_-)$ where $l'_{+/-}\perp l_{+/-}$ (see Fig.~\ref{intersections-degenerate}).
The precise conditions when that happens are as follows. Let us denote the slope of $l_{+/-}$ by $\alpha$. Then, the slope of $l'_{+/-}$ is $-\frac{1}{\alpha}$. Thus, for every $r$ the slope of the lines from \Eref{lines} is either $\alpha$ or $-\frac{1}{\alpha}$, i.e.
\begin{equation}\label{degenerate-condition}
{\textrm{either}}\quad \frac{v^{(n)}(r)_{n-1}}{v^{(n)}(r)_{n}}=\alpha\quad\mathrm{or}\quad\frac{v^{(n)}(r)_{n-1}}{v^{(n)}(r)_{n}}=-\frac{1}{\alpha}\quad{\textrm{for\ all}}\ r=1,\dots,n.
\end{equation}
Define the subsets $\mc{I},\mc{J}\subset\{1,\dots,n\}$, $\mc{I}\cap\mc{J}=\emptyset$ by
\[\mc{I}:=\left\{r:\ \frac{v^{(n)}(r)_{n-1}}{v^{(n)}(r)_{n}}=\alpha\right\},\quad \mc{J}:=\left\{r:\ \frac{v^{(n)}(r)_{n-1}}{v^{(n)}(r)_{n}}=-\frac{1}{\alpha}\right\}.\]
Note that, we can never have $\mc{I}=\emptyset$ or $\mc{J}=\emptyset$, because this implies that the eigenvectors of $A^{(n)}$ are linearly dependent, which is a contradiction. Furthermore, because the eigenvectors of $A^{(n)}$ are orthonormal, we necessarily have
\[\sum_{r=1}^nv^{(n)}(r)_{n-1}v^{(n)}(r)_{n}=0,\quad \sum_{r=1}^n\left(v^{(n)}(r)_{n-1}\right)^2=1,\quad \sum_{r=1}^n\left(v^{(n)}(r)_{n}\right)^2=1.\]
Using $\alpha$, we can express $v^{(n)}_{r,n-1}$ by $v^{(n)}_{r,n}$ and obtain that
\[\alpha S_{\mc{I}}-\frac{1}{\alpha}S_{\mc{J}}=0,\quad \alpha^2 S_{\mc{I}}+\frac{1}{\alpha^2}S_{\mc{J}}=1,\quad S_{\mc{I}}+S_{\mc{J}}=1,\]
where
\[S_{\mc{I}}:=\sum_{i\in\mc{I}}\left(v^{(n)}(i)_{n}\right)^2,\quad S_{\mc{J}}:=\sum_{j\in\mc{J}}\left(v^{(n)}(j)_{n}\right)^2.\]
It is straightforward to see that 
\begin{equation}\label{alpha}
\alpha^2=\frac{S_\mc{J}}{S_\mc{I}}=\frac{1}{S_\mc{I}}-1.
\end{equation}
Finally, note that conditions (\ref{degenerate-condition}) readily imply that the intersecting lines (\ref{lines}) with the circle results with only four points and it is not necessary to look at the free coefficients of the lines. This is because we assume that the spectral data comes from an existing pentadiagonal matrix and hence all lines must intersect in at least one common point.

To sum up, we have the following theorem.
\begin{theorem}\label{thm:penta-main}
For a regular pentadiagonal matrix, the inductive step leads to at most four solutions for $\ket{a^{(n)}}$. Typically, by intersecting the lines from \Eref{lines} there exists only one solution for $\ket{a^{(n)}}$, up to an overall sign. However, there are two possible solutions for $\ket{a^{(n)}}$, up to an overall sign, if and only if the spectral data satisfies the following condition (which we call the $\alpha$-condition): there exists a nonempty subset $\mc{I}\subset\{1,\dots,n\}$, $\mc{I}\neq\{1,\dots,n\}$, such that for all $r=1,\dots,n$ we have $ \frac{v^{(n)}(r)_{n-1}}{v^{(n)}(r)_{n}}=\alpha$ if $r\in\mc{I}$ and $ \frac{v^{(n)}(r)_{n-1}}{v^{(n)}(r)_{n}}=-\frac{1}{\alpha}$ if $r\notin\mc{I}$ for $\alpha$ defined by
\[\alpha=\pm \sqrt{\frac{1}{S_\mc{I}}-1},\quad S_{\mc{I}}:=\sum_{i\in\mc{I}}\left(v^{(n)}(i)_{n}\right)^2.\]
\end{theorem}

\begin{remark}\label{rem-alpha}
The set of matrices satisfying the $\alpha$-condition from Theorem \ref{thm:penta-main} is a subset of all pentadiagonal symmetric matrices of codimension at least one. This explains the results of simulations where sets of a few millions of randomly chosen  pentadiagonal matrices  of dimensions up to $n=10$ where tested, and none satisfied the $\alpha$-condition. This fact can be understood by noting that the matrices satisfying the $\alpha$-condition satisfy certain polynomial equations, as explained in the following Section \ref{subsec:pentaconicsection}. Thus, the set of matrices which do not satisfy the $\alpha$-condition is of a positive codimension and thus it is a full-measure set.
\end{remark}

Another outcome of the above analysis provides the answer to the second question stated at the beginning of the section:  It provides a set of necessary conditions that ensures that the spectral data indeed belong to a pentadiagonal matrix. The conditions are that the distance of each line from \Eref{lines} from the origin must be no greater than $R^{(n)}$, i.e.
\begin{equation}
\frac{ \xi_r^{(n)} }{\sqrt{\left(v^{(n)}(r)_{n-1}\right)^2+\left(v^{(n)}(r)_{n}\right)^2}}\leq R^{(n)}\quad{\textrm{for\ all}}\ r=1,\dots,n.
\end{equation}

\subsubsection{Using the conic sections for recovering pentadiagonal matrices}\label {subsec:pentaconicsection}

\begin{figure}[h]
	\centering
	\includegraphics[width=.8\textwidth]{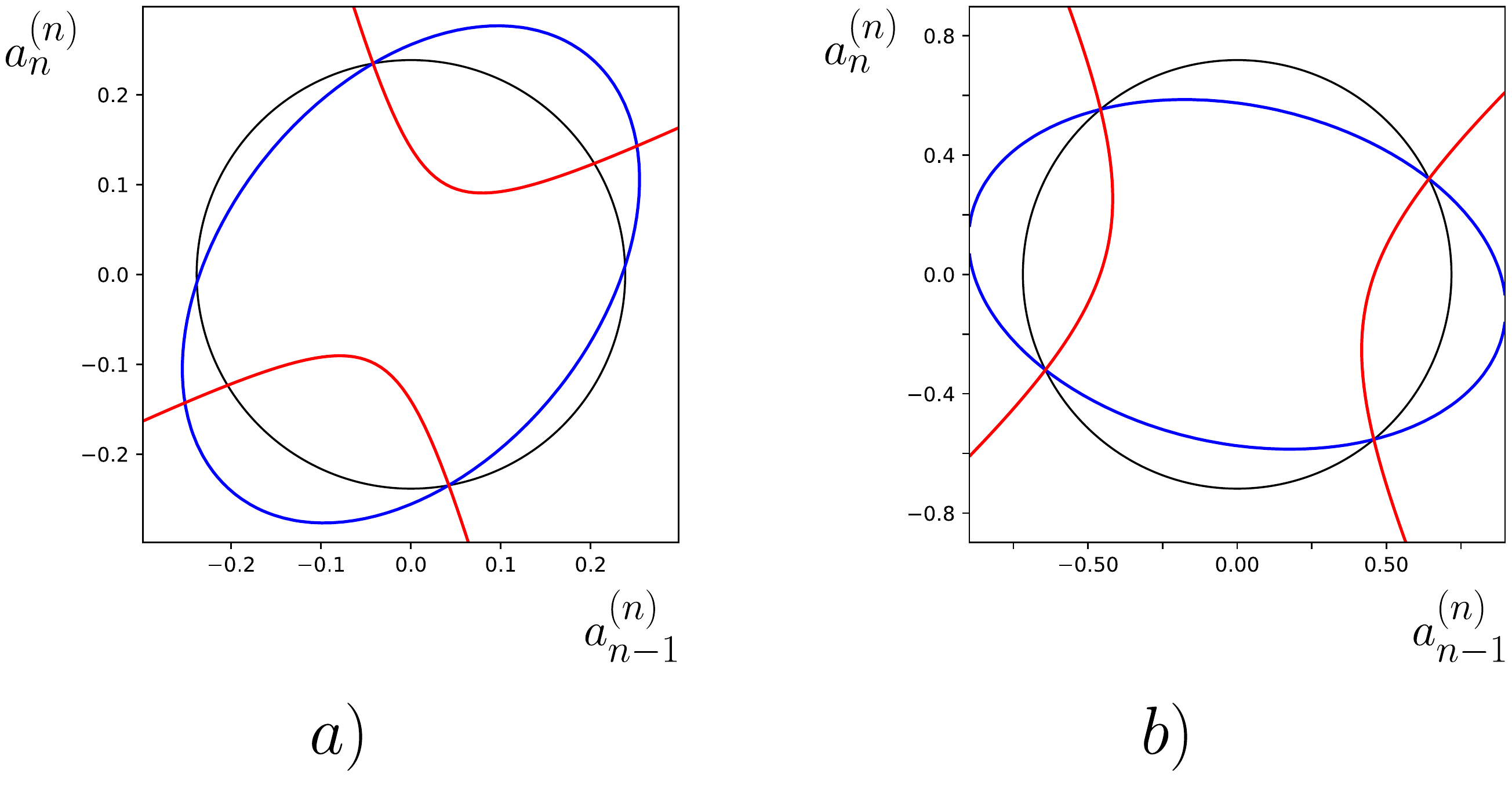}
	\caption{Intersecting the conic sections (\ref{conic1})-(\ref{conic2}) with the circle (\ref{circle}). a) A typical situation where the two conics intersect at two antipodal points which determine $a^{(n)}$ up to an overall sign. b) A degenerate situation when conditions (\ref{penta-degenerate}) are satisfied and the conic sections intersect at four points.}
	\label{fig:conic}
\end{figure}

\begin{figure}[h]
	\centering
	\includegraphics[width=.4\textwidth]{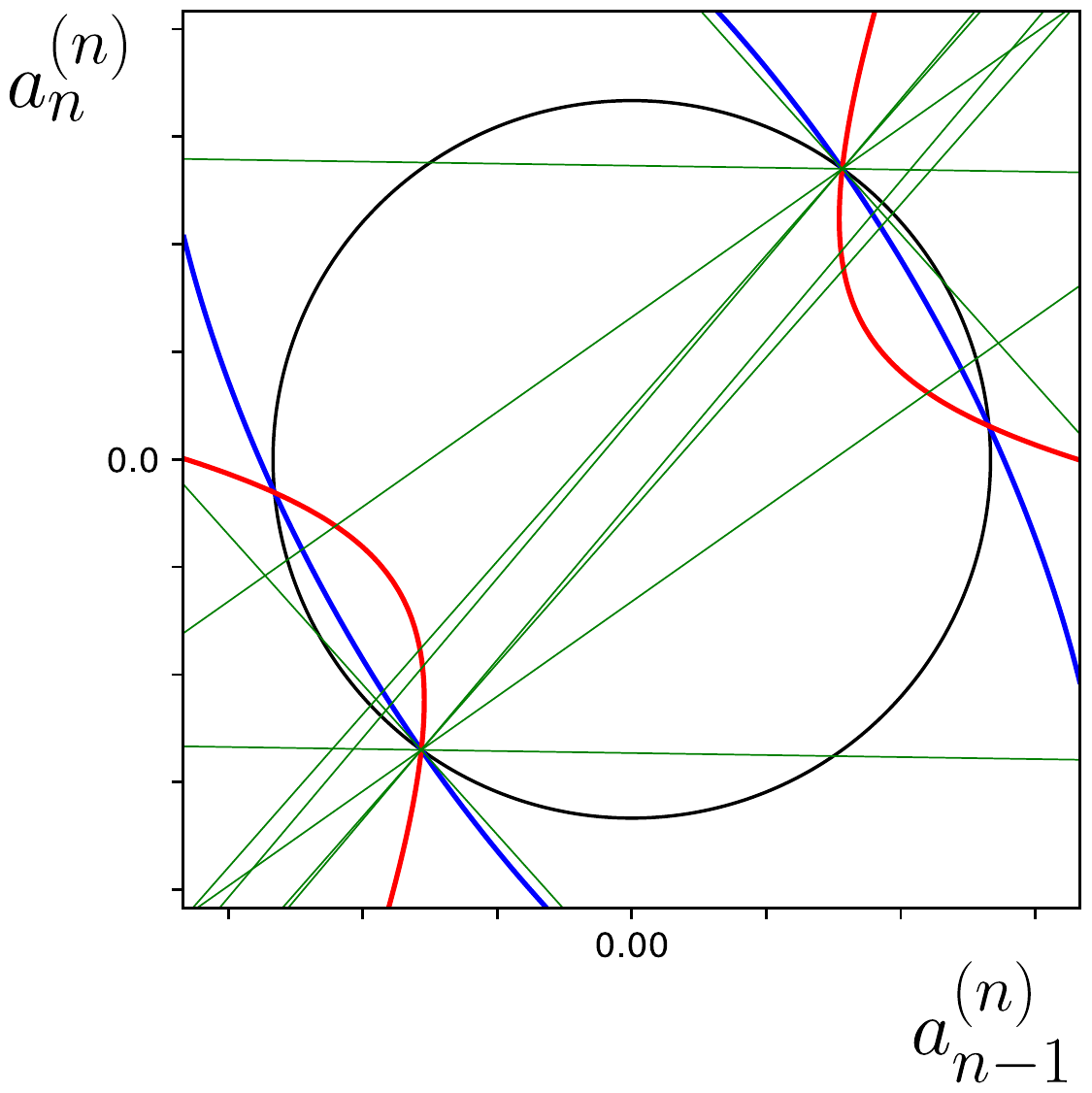}
	\caption{Conic curves (\ref{conic1})-(\ref{conic2}) for a random $5\times 5$ pentadiagonal matrix satisfying degeneracy condition (\ref{penta-degenerate}). Green lines are lines from \Eref{lines} which determine the correct solutions for the column $a^{(5)}$. Despite \Eref{penta-degenerate} being satisfied, only one pair of antipodal points gives solutions for $a^{(5)}$.}
	\label{fig:penta-comparison}
\end{figure}

Before proceeding to the general case, we shall show that using the three quadratic forms from \Eref{radius}, \Eref{third} and \Eref{determinants} typically allow one to reconstruct a pentadiagonal matrix from its spectral data up to an overall sign of each of its columns. For a pentadiagonal matrix, we have
that $\Ket{a^{(n)}}$ has only two non-zero entries and the quadratic forms from \Eref{third} and \Eref{determinants} describe either ellipses or hyperbolae (a.k.a. conic sections). Their explicit forms read
\begin{eqnarray}
\left(a_{n-1}^{(n)}\right)^2+\left(a_{n}^{(n)}\right)^2=\left(R^{(n)}\right)^2, \label{circle}\\
\alpha^{(n)}\left(a_{n-1}^{(n)}\right)^2+2\beta^{(n)}a_{n-1}^{(n)}a_{n}^{(n)}+\gamma^{(n)}\left(a_{n}^{(n)}\right)^2=\rho^{(n)}, \label{conic1}\\
\tilde\alpha^{(n)}\left(a_{n-1}^{(n)}\right)^2+2\tilde\beta^{(n)}a_{n-1}^{(n)}a_{n}^{(n)}+\tilde\gamma^{(n)}\left(a_{n}^{(n)}\right)^2=\tilde\rho^{(n)}, \label{conic2}
\end{eqnarray}
where
\begin{eqnarray*}
	\left(R^{(n)}\right)^2=\frac{1}{2}  \left( \tr\left( (A^{(n+1)})^2\right) - \tr\left( (A^{(n)})^2\right) -h^2  \right), \\
	\rho^{(n)}= \frac{1}{3}  \left( \tr[ (A^{(n+1)})^3] - \tr[ (A^{(n)})^3] -h^3 \right ), \quad \tilde\rho^{(n)}=h-\frac{\det A^{(n+1)}}{\det A^{(n )}}, \\
	\alpha^{(n)}=A^{(n)}_{n-1,n-1}+h,\quad \beta^{(n)}=A^{(n)}_{n-1,n}, \quad \gamma^{(n)}=A^{(n)}_{n,n}+h, \\
	\tilde\alpha^{(n)}=\left(A^{(n)}\right)^{-1}_{n-1,n-1},\quad \tilde\beta^{(n)}=\left(A^{(n)}\right)^{-1}_{n-1,n}, \quad \tilde\gamma^{(n)}=\left(A^{(n)}\right)^{-1}_{n,n}.
\end{eqnarray*}
The three conic sections are centered at $(0,0)$, thus each of the curves from \Eref{conic1} and \Eref{conic2} typically intersects the circle given by \Eref{circle} at exactly four points (or none). There may also be only two intersection points in certain degenerate cases. If any of the intersections is empty or both intersections have zero overlap, then there is no pentadiagonal matrix $A^{(n+1)}$ whose top-left main minor is $A^{(n)}$. If such a pentadiagonal matrix $A^{(n+1)}$ exists, then the three conic sections intersect at at-least two points which yield $a^{(n)}$ up to an overall sign. However, in certain non-generic cases it may happen that the three conic curves intersect at four points (see Fig.~\ref{fig:conic}). 
Then, one cannot decide which of the intersection points give a correct solution. This happens if and only if the principal axes of the two conic sections are identical. This can be written as a condition for the eigenvectors of the quadratic forms defining the conic sections. Namely, both sets of eigenvectors must be the same (up to a scalar factor) \cite{conic}. This in turn happens if and only if the matrices of both quadratic forms commute. This boils down to the following algebraic condition for the coefficients defining the conic intersections (\ref{conic1})-(\ref{conic2}).
\begin{equation}\label{penta-degenerate}
\alpha^{(n)}\tilde\beta^{(n)}-\tilde\alpha^{(n)}\beta^{(n)}+\tilde\gamma^{(n)}\beta^{(n)}-\gamma^{(n)}\tilde\beta^{(n)}=0
\end{equation}
In the special case when $n=2$, \Eref{penta-degenerate} is automatically satisfied and all four points of intersection give correct solutions for $a^{(n)}$. For $n>2$, the set of pentadiagonal $n\times n$ matrices satisfying condition (\ref{penta-degenerate}) is a strict subset of the set of all $n\times n$ pentadiagonal matrices of codimension one. Thus, a typical (generic) pentadiagonal matrix can be completely reconstructed using the above conic curves' method. However, considering non-generic cases brings in a few subtleties. In particular, satisfying the $\alpha$-condition from Theorem \ref{thm:penta-main} implies satisfying \Eref{penta-degenerate}, but the converse statement is not true. The set of $n\times n$ pentadiagonal matrices satisfying the $\alpha$-condition is of codimension at least one, but it is a strict subset of the set of matrices satisfying \Eref{penta-degenerate}. Thus, in non-generic cases only Theorem \ref{thm:penta-main} can give a definite answer about the possible solutions (see Fig.~\ref{fig:penta-comparison}).

\subsection{A finite algorithm for reconstructing a $D$-diagonal matrix.}\label{sub:d-banded}
Given that the spectra $\sigma^{(n)}$, $n=1,\dots,N$ correspond to a $D$-diagonal real symmetric matrix, we provide a finite algorithm which produces the set of possible $D$-diagonal matrices whose main minors have the given spectra $\{\sigma^{(n)}\}_{n=1}^N$. By definition, the $n$th column of a $D$-diagonal matrix satisfies $a^{(n)}_1=\dots=a^{(n)}_{n-d}=0$, where $d:=\frac{1}{2}\left(D-1\right)$. 
 
 The algorithm starts by computing $|\xi_r|=\left|\Braket{a^{(n)}|v^{(n)}(r)}\right|$ from \Eref{xi-equation}. By choosing the sign indicators $\mathbf{s}:=(s_1,\dots,s_n)$, we obtain $2^n$ potential solutions for $\ket{a^{(n)}}$ which read
 \begin{equation}\label{ans}
 \Ket{a^{(n)}(\mathbf{s})}:=\sum_{r=1}^ns_r|\xi_r|\Ket{v^{(n)}(r)}.
 \end{equation}
 For the inductive steps involving the main minors of size $n=1,\dots,d$ these are all valid solutions. However, if $n>d$, we have additional constraints for $\ket{a^{(n)}}$ of the form
 \[\Braket{e^{(n)}(k)|a^{(n)}(\mathbf{s})}=0\quad\mathrm{for}\quad k=1,\dots,n-d.\]
The above equations are satisfied only by some choices of $\mathbf{s}$ which determine the valid solutions in the $n$th inductive step. Alternatively, one can verify the solutions by demanding that vector $\Ket{a^{(n)}(\mathbf{s})}$ belongs to the $(d-1)$-dimensional sphere, i.e.
\begin{equation}\label{sphere-condition}
\sum_{k=n-d}^{n-1}\left|\Braket{e^{(n)}(k)|a^{(n)}(\mathbf{s})}\right|^2=R^{(n)},
\end{equation}
where $R^{(n)}$ depends only on the spectral input data via \Eref{radius}. Thus, we have the following algorithm for realizing the inductive step of constructing $A^{(n+1)}$ from $A^{(n)}$ (Algorithm \ref{step-alg}).
\begin{algorithm}
\caption{Algorithm for constructing $A^{(n+1)}$ from $A^{(n)}$ for a $D$-banded matrix.}\label{step-alg}
\begin{algorithmic}[1]
\State \textbf{Input:} \text{$A^{(n)}$ and its spectral decomposition, $\sigma^{(n+1)}$, $\epsilon$ - numerical accuracy.}
\State \textbf{Output:} \text{The main minor $A^{(n+1)}$.}
\State $d=\frac{1}{2}\left(D-1\right) \gets \text{size of the new column}$
\State $\{|\xi_{r}^{(n)}|\}_{r=1}^n \gets \text{list of overlaps from \Eref{xi-equation}}$
\State $solutions \gets \text{list of solutions for the vector $a^{(n)}$}$
\For{${\bf{s}}\in\{0,1\}^n$}
\State $\Ket{a^{(n)}(\mathbf{s})}\gets \text{vector from \Eref{ans}}$
\If{$\left|\sum_{k=n-d+1}^{n}\left|\Braket{e^{(n)}(k)|a^{(n)}(\mathbf{s})}\right|^2-R^{(n)}\right|\leq\epsilon$}
\State $solutions$.append$\left(\Ket{a^{(n)}(\mathbf{s})}\right)$
\EndIf
\EndFor
\State \textbf{return} $solutions$
\end{algorithmic}
\end{algorithm}
One can see that the main contribution to the numerical complexity of the above inductive step relies on the loop going through all $2^n$ choices of the possible signs. Thus, the complexity scales like $poly_{n,d}2^n$, where the polynomial factor $poly_{n,d}$ comes from the necessity of evaluating expressions from \Eref{ans} and \Eref{sphere-condition} in each step of the loop. What is more, heuristics shows that the required numerical accuracy $\epsilon$ grows with $d$ and $n$, as the potential solutions tend to bunch close to the surface of the sphere. 

\begin{figure}[h]
	\centering
	\includegraphics[width=.5\textwidth]{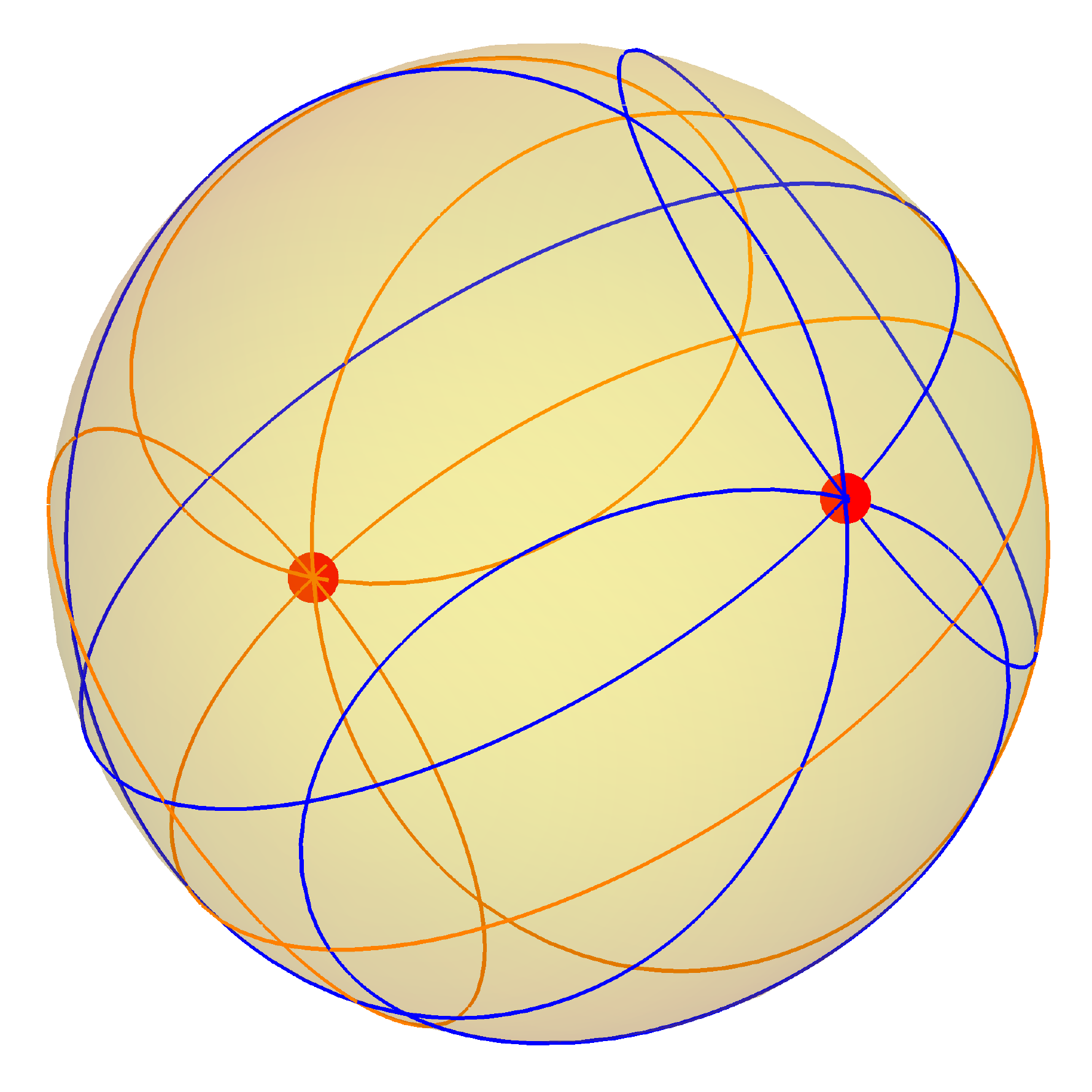}
	\caption{The result of intersecting eight hyperplanes $H_r^{(n),\pm}$, $r=1,2,3,4$ for a typical random heptadiagonal matrix. Intersecting one of the hyperplanes with the sphere of radius $R^{(n)}$ results with a single circle. The common points where $n$ circles meet (here $n=4$) determine the solutions for the vector $a^{(n)}$ (the red dots). As one can see, typically there are just two such points which lie on the opposite sides of the sphere. The orange and blue colors were used to mark the relevant groups of intersecting circles.}
	\label{hepta}
\end{figure}

Let us next argue that for generic input data, the result of the above Algorithm \ref{step-alg} for $\epsilon$ sufficiently small are just two antipodal solutions for $a^{(n)}$. To this end, consider the hyperplanes
\[H_r^{(n),\pm}:\quad a_{n-d+1}v^{(n)}(r)_{n-d+1}+\dots+a_{n}v^{(n)}(r)_{n}=\pm|\xi_r|.\]
The procedure of intersecting lines with the circle from Subsection \ref{subsec:penta-main-minors} for $d>2$ generalizes here to a procedure involving the study of intersections of hyperplanes $H_r^{(n),\pm}$ with the (hyper)sphere of radius $R^{(n)}$. As before, typically, as a result of such a procedure one obtains only a single pair of antipodal solutions for $a^{(n)}$. Non-typical situations arise when some hyperplanes are perpendicular to each other. This requires the existence of certain relations between the eigenvectors of $A^{(n)}$ analogously to the relations stated in Theorem \ref{thm:penta-main}. However, the precise form of these conditions seems to be much more complicated and we leave this problem open for future research. For an illustration of the intersection procedure for a typical heptadiagonal matrix ($d=3$), see Fig.~\ref{hepta}.

Similarly to the pentadiagonal case, one can find necessary conditions for the spectral data to correspond to a $D$-diagonal matrix. Every hyperplane $H_r^{(n),\pm}$ must have a nontrivial intersection with the sphere of radius $R^{(n)}$, hence its distance from the origin must be no greater than $R^{(n)}$.
\[\frac{|\xi_r|}{\sqrt{\sum_{k=1}^d\left(v^{(n)}(r)_{n-d+k}\right)^2}}\leq R^{(n)}\quad{\textrm{for\ all}}\ r=1,\dots,n.\]

\begin{remark}
Using Theorem \ref{thm:degenerate} we can apply the above methodology mutatis mutandis to non-regular banded matrices. In particular, whenever the dimension of the smaller truncated matrix from Theorem \ref{thm:degenerate} is greater than $d$, we can use the same redundancy effects to our advantage when dealing with typical matrices.
\end{remark}

\section{Improved strategies for $D$-diagonal matrices -- the sliding minor construction}\label{sec:optimal}
The number of independent non vanishing entries of a $D$-diagonal real symmetric matrix is
\begin{equation}
N_D=\frac{1}{2}(2N-d)(d+1), \ \  {\rm where}\ \  D=2d+1 .
\end{equation} 
The spectral data for all the main minors of a $D$-diagonal  $N\times N$ matrix is $N_{N-1}$, which exceeds $N_D$ by far. In this section, we provide two  alternative methods for reconstructing $D$-diagonal matrices and which use  much less spectral data. The first  utilizes the minimum  number of spectral parameters needed for the purpose, together with the necessary sign indicators. The second method makes use of $N-d-1$ more spectral data. This introduces enough redundancy to render the method much more straight forward to use at the cost of being applicable only for {\it  generic} matrices.  

\subsection{Inverse method  with minimal spectral input}
The  number of non vanishing  data needed to write a $D$-diagonal matrix,  $N_D$, can also be written as $ N_{D}=\frac{1}{2}d(d+1)+(d+1)(N-d)$. It can be interpreted as the sum of two terms:
\begin{enumerate}
\item The spectra of the minors  $A^{(1)}=A_{1,1}$, $A^{(2)}$, $\dots$, $A^{(d)}$ which give a total of $\frac{1}{2}d(d+1)$ numbers needed to reconstruct the first $d\times d$ minor.  
\item  The spectra of $(N-d)$ minors of size $(d+1)\times (d+1)$ with upper diagonal entry at successive positions along the diagonal of $A$ which are denoted by  $M_1^{(d+1)}=A^{(d+1)}$, $M_2^{(d+1)}$, $\dots$, $M_{N-d}^{(d+1)}$ (see (\ref{frames}) for an example with $N=6$ and $d=2$). The "sliding minors" in the present construction are the 
 $\left \{ M_k^{(d+1)}\right \}_{k=2}^{N-d}$ minors. 
\begin{equation}\label{frames}
  \tikz[baseline=(M.west)]{%
    \node[matrix of math nodes,matrix anchor=west,left delimiter=(,right delimiter=),ampersand replacement=\&] (M) {%
      A_{1,1} \& \star \& \star \& 0 \& 0 \& 0 \\
      \star \& A_{2,2} \& \star \& \star \& 0 \& 0 \\
      \star \& \star \& A_{3,3} \& \star \& \star \& 0 \\
      0 \& \star \& \star \& A_{4,4} \& \star \& \star \\
      0 \& 0 \& \star \& \star \& A_{5,5} \& \star \\
      0 \& 0 \& 0 \& \star \& \star \& A_{6,6} \\
    };
    \node[draw,fit=(M-1-1)(M-3-3),inner sep=-1pt] {};
    \node[draw,fit=(M-2-2)(M-4-4),inner sep=-1pt] {};
    \node[draw,fit=(M-3-3)(M-5-5),inner sep=-1pt] {};
    \node[draw,fit=(M-4-4)(M-6-6),inner sep=-1pt] {};
  }
\end{equation}
\end{enumerate}

\noindent  The reconstruction algorithm consists of the following steps. 
\begin{enumerate}[label=(\arabic*)] 
\item Reconstruct  $A^{(d+1)}$ from the spectral data of $A^{(1)}$ through $A^{(d)}$ and the sign indicators ($\frac{1}{2}d(d+1)$ signs in total) as described in the inductive procedure from the proof of Theorem \ref{thm:main}.

\item The matrix elements of the $d\times d$ minor $M_2^{(d)}$ starting with the upper diagonal $A_{2,2}$ are obtained from the known spectral decomposition of  $A^{(d+1)}=M_1^{(d+1)}$. The eigenvalues and eigenvectors of $M_2^{(d)}$ are then computed.    

\item Use Theorem \ref{thm:main} to reconstruct $M_2^{(d+1)}$ from the spectral decomposition of $M_2^{(d)}$ and appropriate sign indicator data.

\item Repeat the previous two steps recursively  to procedure the successive minors $M_{3}^{(d+1)},\cdots, M_{N-d}^{(d+1)}$.
\end{enumerate}

The main drawback of this method is the need to provide the sign indicators at each step. To overcome this,    
one can also introduce minimal redundancy to the above sliding-minor method so that for a typical matrix the required sign data reduces only to the overall signs of columns of $A^{(N)}$. This applies only for generic matrices, as explained in the previous section. 
 
\subsection{Inverse method  with optimal spectral input} 
Here, the sliding minors are the  $\left \{ M_k^{(d+2)}\right \}_{k=2}^{N-d-1}$ minors (see (\ref {bigframes})).   
The increase of their dimension amounts also to the vanishing of  the  $(1,d+2)$ and  $(d+2,1)$ entries.  This   enables turning  the redundant information to impose  constraints, which for generic matrices remove  the need for computing the sign -indicators required in the previous construction. 
\begin{equation}\label{bigframes}
\tikz[baseline=(M.west)]{%
	\node[matrix of math nodes,matrix anchor=west,left delimiter=(,right delimiter=),ampersand replacement=\&] (M) {%
		A_{1,1} \& \star \& \star \& 0 \& 0 \& 0 \\
		\star \& A_{2,2} \& \star \& \star \& 0 \& 0 \\
		\star \& \star \& A_{3,3} \& \star \& \star \& 0 \\
		0 \& \star \& \star \& A_{4,4} \& \star \& \star \\
		0 \& 0 \& \star \& \star \& A_{5,5} \& \star \\
		0 \& 0 \& 0 \& \star \& \star \& A_{6,6} \\
	};
	\node[draw,fit=(M-1-1)(M-4-4),inner sep=-1pt] {};
	\node[draw,fit=(M-2-2)(M-5-5),inner sep=-1pt] {};
	\node[draw,fit=(M-3-3)(M-6-6),inner sep=-1pt] {};
}
\end{equation}
\noindent  The inversion algorithm consists of the following steps. 
\begin{enumerate}[label=(\arabic*)] 
\item Reconstruct  $A^{(d+1)}$ from the spectral data of $A^{(1)}$ through $A^{(d)}$ and the sign indicators ($\frac{1}{2}d(d+1)$ signs in total) as described in the inductive procedure from the proof of Theorem \ref{thm:main}.
\item The matrix elements of the $(d+2)\times (d+2)$ minor $M_1^{(d+2)}$ starting at the  upper diagonal $A_{1,1}$ are computed using the algorithm proposed in section (\ref{sub:d-banded}). The only sign indicator needed is the sign of the entry $A_{1,d+1}$.
\item The $(d+1) ,(d+1) $ minor $M_2^{(d+1)}$ is extracted  from the computed  $M_1^{(d+2)}$.     Its eigenvalues and eigenvectors  are then computed.  ($M_2^{(d+1)}$ is the $d+1\times d+1$ minor with $A_{2,2}$ as its upper diagonal entry)  
\item Use Theorem \ref{thm:main} to reconstruct $M_3^{(d+2)}$ from the spectral decomposition of $M_2^{(d+1)}$ and the sign of the entry $A_{1,d+1}$.
\item Repeat the previous two steps recursively  to produce the successive minors $M_{3}^{(d+1)},\cdots, M_{N-d-1}^{(d+2)}$. 
\end{enumerate}

In order to minimize further the set of exceptional matrices where this strategy fails, one could choose a larger dimension for the sliding minor. In choosing an "optimum" strategy, one has to weigh computational effort against minimizing the exceptional set. This is an individual decision left for the discretion of the practitioner.    

\subsection{Pentadiagonal matrices}
The first inversion strategy for regular pentadiagonal matrices ($d=2$) is a combination of two steps which we have already analyzed in detail in the previous sections. The first step of reconstructing $A^{(2)}$ has been described in the proof of Theorem \ref{thm:main} (see \Eref{a2} and \Eref{a2-eigenv}). The second as well as all other steps rely on reconstructing $3\times 3$ matrices $\left \{M_k^{(3)}\right \}_{k=2}^{N-2}$ from their top-left $2\times 2$ minors. This has been done explicitly in subsection \ref{3x3-example}. In total, the $2\times 2\to 3\times 3$ steps require the following sign data.
\begin{enumerate}
\item The sign of the matrix element $A_{1,2}$.
\item A pair of signs from \Eref{signs2to3} applied to each minor $\left \{M_k^{(3)}\right \}_{k=2}^{N-2}$.
\end{enumerate}

The second inversion strategy which  uses the larger $4\times 4$ sliding minor gives extra redundancy and, as explained above, it typically reduces the required sign data to the overall signs of columns of $A^{(N)}$. Here, the notion of typicality is particularly straightforward to state. Namely, all the sliding minors of a typical matrix do not satisfy the $\alpha$-condition from Theorem \ref{thm:penta-main}\ . The computation of the spectral decomposition of the $(3\times 3)$  minors can be carried out analytically. 

\section{Summary}
We revisited here the question of how much spectral data is needed to determine a real symmetric matrix. In principle, a matrix is determined by its spectrum only up to unitary equivalence. Hence, more spectral information is needed in order to reconstruct a non-diagonal matrix. Heuristically, the number of spectral data should match the number of unknowns. Hence, for a generic real symmetric $N\times N$ matrix, we need $N(N+1)/2$ numbers coming from the spectral data. In our work, these numbers come from the spectra of the main minors. Given such  spectra, we provide a finite algorithm that reconstructs the matrix up to a finite number of possibilities. The construction can be made unique by supplementing additional sign data as stated in Theorem \ref{thm:main} which comprises of $N(N-1)/2$ signs in total. However, the signs are not easily deduced from the matrix itself, therefore the requirement of supplementing the signs is a notable limitation to our procedure. Thus, in further parts of this paper, we focus on ways of reducing the required sign data in the case of banded matrices. For $(2d+1)$-banded matrices, there are less unknowns, hence one can find a smaller optimal set of $(d+1)\times (d+1)$ minors whose spectra allow one to reconstruct the matrix up to a finite number of possibilities. As before, some additional sign information is required to make the construction unique. By analysing the case of pentadiagonal matrices, we compare both methods of reconstruction. We find that the redundant information coming from the spectra of all main minors in a typical case (where the notion of {\it typicality} has been made precise in Subsection \ref{subsec:penta-main-minors}) allows one to reduce the number of required signs to $N-1$ that are just the overall signs of columns of the upper-triangular part of $A^{(N)}$, while using the optimal amount of spectral information requires $2N-1$ signs. In Section \ref{sec:optimal} we argue that this is a general fact, i.e. the redundancy coming from all main minors typically allows one to uniquely reproduce a general banded matrix from the spectral data using only the overall signs of columns of the upper-triangular part of $A^{(N)}$.

\section *{Acknowledgment}
We would like to thank Professor Raphael Loewy for suggestions and critical comments during the first stages of the present work. Professor Percy Deift accompanied the work since its beginning and offered critique and advise in many instances. We are obliged for his invaluable help and in particular for his permission to quote an unpublished theorem quoted in Remark \ref {per1}. 
Professor Carlos Meite is acknowledged for bringing to our attention his work (Remark \ref {carlos}) before its publication.  We also thank Professor Jon Keating for useful suggestions and Doctor Yotam Smilansky for helpful comments on the manuscript. The inspiration for the work came by listening to a seminar by Professor Christiane Tretter within the webinar series "Spectral Geometry in the Clouds".  Thanks Christiane. The seminar is organized by Doctors Jean Lagacé and Alexandre Girouard  whom we thank for initiating and running such successful  seminar during the miserable Corona days. 

 \appendix
 \section{Identities for Cauchy Matrices}\label{appA}
 
\noindent For two sequences of real numbers, $\bx:=\{x_i\}_{i=1}^n$ and $\by:=\{y_i\}_{i=1}^n$, define the Cauchy matrix $C(\bx,\by)$ as $\left(C(\bx,\by)\right)_{i,j}:=\frac{1}{x_i-y_j}$. The matrix $C(\bx,\by)$ has the followng symmetry
\begin{equation}\label{cauchy-symmetry}
\left(C(\bx,\by)\right)^T=-C(\by,\bx).
\end{equation}
$C$ will be used to denote $C(\bx,\by)$ by default. An explicit expression for $C^{-1}$ reads \cite {Gow 1992}
\begin{equation} 
\hspace{-10mm}(C^{-1})_{i,j}=\left [(y_i-x_i)  \prod_{k\ne i}^n\frac{y_i-x_k}{y_i-y_k}\right ] \frac{1}{y_i-x_j} \left[(x_j-y_j) \prod_{k\ne j}^n\frac{x_j-y_k}{x_j-x_k} \right ]\  
\end{equation}
The sum of the matrix elements of $C^{-1}$ is known to be  \cite {ident}
\begin{equation}
\sum_{i,j=1}^n (C^{-1})_{i,j} \ =\sum_{k=1}^n (x_k-y_k)
\end{equation}

\noindent We aim to prove the following identities
\begin{equation}\label{cauchy-identities1}
\sum_{i,j}\left(C^{-1}\right)_{i,j}\frac{1}{x_j}=1-\frac{\prod_{i=1}^ny_i}{\prod_{i=1}^nx_i},\quad \sum_{i,j}\frac{1}{y_i}\left(C^{-1}\right)_{i,j}=\frac{\prod_{i=1}^nx_i}{\prod_{i=1}^ny_i}-1,
\end{equation}

\begin{equation}\label{cauchy-identities2}
\fl \sum_{i,j}y_i\left(C^{-1}\right)_{i,j}\frac{1}{x_j}=\frac{\prod_{i=1}^ny_i}{\prod_{i=1}^nx_i}\sum_{j=1}^n\left(x_j-y_j\right),\quad \sum_{i,j}\frac{1}{y_i}\left(C^{-1}\right)_{i,j}x_j=\frac{\prod_{i=1}^nx_i}{\prod_{i=1}^ny_i}\sum_{j=1}^n\left(x_j-y_j\right).
\end{equation}

\noindent Note first, that the LHS identities imply the RHS identities by symmetry \ref{cauchy-symmetry}. Hence, it is enough to prove the LHS identities. To this end, we write them down as the following matrix-vector products
\begin{equation}\label{vec-prod}
\fl \sum_{i,j}\left(C^{-1}\right)_{i,j}\frac{1}{x_j}=\Bra{1^{(n)}} C^{-1}\ket{\frac{1}{x}^{(n)}},\quad \sum_{i,j}y_i\left(C^{-1}\right)_{i,j}\frac{1}{x_j}=\Bra{y^{(n)}} C^{-1}\ket{\frac{1}{x}^{(n)}},
\end{equation}
where vectors $\ket{1^{(n)}}$, $\ket{\frac{1}{x}^{(n)}}$ and $\ket{y^{(n)}}$ are defined as
\begin{equation*}
\ket{1^{(n)}}:=
\left(\begin{array}{l}
1 \\
1 \\
\vdots \\
1
\end{array}\right),\quad
\ket{\frac{1}{x}^{(n)}}:=
\pmatrix{
1/x_1 \cr
1/x_2 \cr
\vdots \cr
1/x_n
},\quad
\ket{y^{(n)}}:=
\left(\begin{array}{l}
y_1 \\
y_2 \\
\vdots \\
y_n
\end{array}\right).
\end{equation*}
We will use the following lemma.
\begin{lemma}[Vandermonde Matrix Identity for Cauchy Matrix]\label{vandermonde}
$C(\bx,\by)$ has the following decomposition
\[C(\bx,\by)=-PV_x^{-1}V_yQ^{-1},\]
where $P$ and $Q$ are diagonal matrices defined as
\[P={\mathrm{diag}}\left(p_1(x_1),\dots,p_n(x_n)\right),\quad Q={\mathrm{diag}}\left(p(y_1),\dots,p(y_n)\right)\]
with
\[p(t):=\prod_{i=1}^n(t-x_i),\quad p_k(t):=\prod_{i=1,i\neq k}^n(t-x_i),\quad 1\leq k\leq n,\]
and $V_x$, $V_y$ are the Vandermonde matrices
\begin{equation*}
V_x=
\left(\begin{array}{cccc}
1 & 1 &  \cdots & 1 \\ 
x_1 & x_2 &  \cdots & x_n \\
\vdots & \vdots  & \ddots &\vdots \\
x_1^{n-1} & x_2^{n-1} & \cdots & x_n^{n-1}
\end{array}\right),\quad 
V_y=
\left(\begin{array}{cccc}
1 & 1 &  \cdots & 1 \\ 
y_1 & y_2 &  \cdots & y_n \\
\vdots & \vdots  & \ddots &\vdots \\
y_1^{n-1} & y_2^{n-1} & \cdots & y_n^{n-1}
\end{array}\right).
\end{equation*}
\end{lemma}
\noindent By Lemma \ref{vandermonde}, we have
\[C^{-1}=-QV_Y^{-1}V_xP^{-1}.\]
Next, we define the vectors $\ket{\phi_1}$, $\ket{\phi_y}$ and $\ket{\phi_x}$ as
\begin{equation}\label{vecs-def}
\fl \Bra{\phi_1}:=-\Bra{1^{(n)}}QV_Y^{-1},\quad \Bra{\phi_y}:=-\Bra{y^{(n)}}QV_Y^{-1},\quad \ket{\phi_x}:=V_xP^{-1}\ket{\frac{1}{x}^{(n)}}.
\end{equation}
Calculation of \ref{vec-prod} boils down to computing overlaps
\begin{equation*}
	\sum_{i,j}\left(C^{-1}\right)_{i,j}\frac{1}{x_j}=\Braket{\phi_1|\phi_x},\quad \sum_{i,j}y_i\left(C^{-1}\right)_{i,j}\frac{1}{x_j}=\Braket{\phi_y|\phi_x}.
\end{equation*}
Let us start with finding $\ket{\phi_x}$. A straightforward calculation using \Eref{vecs-def} shows
\begin{equation}\label{ux-sums}
\ket{\phi_x}=
\left(\begin{array}{l}
\sum_{k=1}^n\frac{1}{x_kp_k(x_k)} \\
\sum_{k=1}^n\frac{1}{p_k(x_k)} \\
\vdots \\
\sum_{k=1}^n\frac{x_k^{n-2}}{p_k(x_k)}
\end{array}\right).
\end{equation}
In order to compute the sums in \Eref{ux-sums}, we use the following Lemma \cite{knuth}.
\begin{lemma}[Summation of Powers over Product of Differences]\label{summation}
Let $(t_1,t_2,\dots,t_n)$ be a sequence of real numbers. Then,
\begin{equation*}
\sum_{k=1}^n\left(\frac{t_k^m}{\prod_{i=1,i\neq k}(t_k-t_i)}\right)=
\cases{
0 & $: 0 \le m < n - 1$ \cr
1 & $: m = n - 1$ \cr
\sum_{j = 1}^n t_j & $: m = n$
}
\end{equation*}
\end{lemma}
\noindent By Lemma \ref{summation}, we have $\Braket{e_j^{(n)}|\phi_x}=$ for $j\geq 2$. The only nonzero entry is 
\begin{equation}\label{phix}
\Braket{e_1^{(n)}|\phi_x}=\sum_{k=1}^n\frac{1}{x_kp_k(x_k)}=-\frac{(-1)^{n}}{x_1\cdotp\dots\cdotp x_n},
\end{equation}
which is evaluated in the following Proposition. 
\begin{prop}\label{summation1}
	Let $(t_1,t_2,\dots,t_n)$ be a sequence of real numbers. Then,
	\[\sum_{k=1}^n\frac{1}{t_k\prod_{i\neq k}(t_k-t_i)}=-\frac{(-1)^{n}}{t_1\cdotp\dots\cdotp t_n}.\]
\end{prop}
\begin{proof}
	Consider the following complex integral over a circle of radius $R>\max\{|t_i|:\ 1\leq i\leq n\}$
	\begin{equation*}
		I_R(t_1,\dots,t_n):=\frac{1}{2\pi i}\oint_{|z|=R}\frac{dz}{z(z-t_1)\cdotp\dots\cdotp(z-t_n)}.
	\end{equation*}
	Because the integrand is a regular function at $z=\infty$, we have $I_R(t_1,\dots,t_n)=0$. On the other hand, the integrand has simple poles at $z=0$, $z=t_1$, $\dots$, $z=t_n$. Hence, the residue theorem asserts that
	\begin{equation*}
		I_R(t_1,\dots,t_n)=(-1)^n\frac{1}{t_1\cdotp\dots\cdotp t_n}+\sum_{k=1}^n\frac{1}{t_k\prod_{i\neq k}(t_k-t_i)},
	\end{equation*}
	which finishes the proof.
\end{proof}

Next, let us compute the vectors $\ket{\phi_1}$ and $\ket{\phi_y}$. Note first that because the first entry of $\ket{\phi_x}$ is its only nonzero entry, we only need to find the first entries of $\ket{\phi_1}$ and $\ket{\phi_y}$. By \Eref{vecs-def}, the desired vectors are solutions to the following linear equations
\[V_Y^T\ket{\phi_1}=-Q^T\ket{1^{(n)}},\quad V_Y^T\ket{\phi_y}=-Q^T\ket{y^{(n)}}.\]

\begin{equation*}
	Q^T\ket{1^{(n)}}=
	\left(\begin{array}{l}
		p(y_1) \\
		p(y_2) \\
		\vdots \\
		p(y_3)
	\end{array}\right),\quad
	Q^T\ket{y^{(n)}}=
	\left(\begin{array}{l}
		y_1p(y_1) \\
		y_2p(y_2) \\
		\vdots \\
		y_n p(y_n)
	\end{array}\right).
\end{equation*}

\noindent  We use now Crammer's rule.
\begin{equation*}
	\Braket{e_1^{(n)}|\phi_1}=-\frac{1}{\det V_y^T}
	\left(\begin{array}{ccccc}
		p(y_1) & y_1 & y_1^2 & \cdots & y_1^{n-1} \\ 
		p(y_2) & y_2 & y_2^2 & \cdots & y_2^{n-1} \\
		\vdots & \vdots & \vdots & \ddots &\vdots \\
		p(y_n) & y_n & y_n^2 & \cdots & y_n^{n-1} \\ 
	\end{array}\right)
\end{equation*}

\begin{equation*}
	\Braket{e_1^{(n)}|\phi_y}=-\frac{1}{\det V_y^T}
	\left(\begin{array}{ccccc}
		y_1p(y_1) & y_1 & y_1^2 & \cdots & y_1^{n-1} \\ 
		y_2p(y_2) & y_2 & y_2^2 & \cdots & y_2^{n-1} \\
		\vdots & \vdots & \vdots & \ddots &\vdots \\
		y_np(y_n) & y_n & y_n^2 & \cdots & y_n^{n-1} \\ 
	\end{array}\right) \ .
\end{equation*}

\noindent Applying  the Laplace expansion of both determinants with respect to their first columns.
\begin{eqnarray*}
	\Braket{e_1^{(n)}|\phi_1}=-\frac{1}{\det V_y^T}\sum_{j=1}^n (-1)^j p(y_j)\det \left(V\left(y_1,\dots,\hat y_j,\dots,y_n\right)\right)^T, \\
	\Braket{e_1^{(n)}|\phi_y}=-\frac{1}{\det V_y^T}\sum_{j=1}^n (-1)^j y_jp(y_j)\det  \left(V\left(y_1,\dots,\hat y_j,\dots,y_n\right)\right)^T,
\end{eqnarray*}
where
\begin{equation*}
	\left(V\left(y_1,\dots,\hat y_j,\dots,y_n\right)\right)^T:=
	\left(\begin{array}{cccc}
		y_1 & y_1^2 & \cdots & y_1^{n-1} \\ 
		\vdots & \vdots & \ddots &\vdots \\
		y_{j-1} & y_{j-1}^2 & \cdots & y_{j-1}^{n-1} \\
		y_{j+1} & y_{j+1}^2 & \cdots & y_{j+1}^{n-1} \\
		\vdots & \vdots & \ddots &\vdots \\
		y_n & y_n^2 & \cdots & y_n^{n-1} \\ 
	\end{array}\right)
\end{equation*}
Using the expression for the determinant of a Vandermonde matrix, we have 
\[\det \left(V\left(y_1,\dots,\hat y_j,\dots,y_n\right)\right)^T=\left(\prod_{k\neq j}y_k\right)\cdotp\left(\prod_{k<l,k\neq j,l\neq j}(y_l-y_k)\right)\]
It follows that
\[\frac{\det \left(V\left(y_1,\dots,\hat y_j,\dots,y_n\right)\right)^T}{\det V_y^T}=(-1)^{n-j}\frac{\prod_{k\neq j}y_k}{\prod_{k\neq j}(y_k-y_j)}=(-1)^{n-j}\frac{\prod_{k\neq j}y_k}{p_j(y_j)}.\]
After application of the above result, the desired expressions read
\begin{equation}\label{vecs-intermediate}
\fl \Braket{e_1^{(n)}|\phi_1}=(-1)^ny_1\cdotp\dots\cdotp y_n\sum_{j=1}^n \frac{p(y_j)}{y_jp_j(y_j)}, \quad \Braket{e_1^{(n)}|\phi_y}=(-1)^ny_1\cdotp\dots\cdotp y_n\sum_{j=1}^n \frac{p(y_j)}{p_j(y_j)}.
\end{equation}
As the final step, we use the following expansion of $p(t)$.
\begin{equation*}
	p(t)=t^n+\sum_{l=1}^na_lt^{l-1},\quad a_l=(-1)^{n+1-l}e_{n+1-l}\left(\{x_1,\dots,x_n\}\right),
\end{equation*}
where $e_{n+1-l}\left(\{x_1,\dots,x_n\}\right)$ is the $(n+1-l)$th Viete sum in variables $\{x_1,\dots,x_n\}$. Plugging the above expansion into \Eref{vecs-intermediate}, we get
\begin{equation*}
	\Braket{e_1^{(n)}|\phi_1}=(-1)^ny_1\cdotp\dots\cdotp y_n\left(\sum_{j=1}^n\frac{y_j^{n-1}}{p_j(y_j)}+\sum_{l=1}^na_l\sum_{j=1}^n\frac{y_j^{l-2}}{p_j(y_j)}\right).
\end{equation*}
\begin{equation*}
	\Braket{e_1^{(n)}|\phi_y}=(-1)^ny_1\cdotp\dots\cdotp y_n\left(\sum_{j=1}^n\frac{y_j^{n}}{p_j(y_j)}+\sum_{l=1}^na_l\sum_{j=1}^n\frac{y_j^{l-1}}{p_j(y_j)}\right).
\end{equation*}
By Lemma \ref{summation}, we get that
\begin{eqnarray*}
	\Braket{e_1^{(n)}|\phi_1}=(-1)^ny_1\cdotp\dots\cdotp y_n\left(1+a_1\sum_{j=1}^n\frac{1}{y_jp_j(y_j)}\right), \\
	\Braket{e_1^{(n)}|\phi_y}=(-1)^ny_1\cdotp\dots\cdotp y_n\left(\sum_{j=1}^ny_j+a_n\right).
\end{eqnarray*}
Finally, after making use of Proposition \ref{summation1} and expanding Viete sums $a_1=(-1)^n x_1\cdotp\dots\cdotp x_n$ and $a_n=-(x_1+\dots+x_n)$
\begin{equation*}
\fl	\Braket{e_1^{(n)}|\phi_1}=(-1)^n(y_1\cdotp\dots\cdotp y_n-x_1\cdotp\dots\cdotp x_n), \quad
	\Braket{e_1^{(n)}|\phi_y}=(-1)^ny_1\cdotp\dots\cdotp y_n\sum_{j=1}^n(y_j-x_j).
\end{equation*}
The result follows now directly by multiplying the above expressions by $\Braket{e_1^{(n)}|\phi_x}$ from \Eref{phix}.

The identity which is used in the discussion of  (\ref {consistent}) follows by combining \Eref{cauchy-identities1} and \Eref{cauchy-identities2} to give 
\begin{equation}
s=\sum_{i,j}\frac{1}{y_i} (C^{-1} )_{i,j} (x_j-s)\ , {\rm where} \ s=\sum_{k=1}^n (x_k-y_k)\ .
\end{equation}

 \section{Derivation of the formula for $({\xi_r}^{(n)}) ^2$.}\label{appB}
 The aim is to prove the identity used in \Eref{xi-equation} which reads
 \begin{equation}\label{identity-xi}
\sum_{k=1}^n \left (  C^{-1}\right )_{r,k}\  (   x_k -h)=-\frac{\prod_{k=1}^{n+1}\left(y_r-x_k\right)}{\prod_{k=1,k\neq r}^n\left(y_r-y_k\right)}, 
\end{equation}
where $C=C(\bx,\by)$ is a Cauchy matrix and $h:=\sum_{k=1}^{n+1}x_k-\sum_{k=1}^ny_n$.
Note first that 
\begin{equation}\label{sums}
\fl \sum_{k=1}^n \left (  C^{-1}\right )_{r,k}\     x_k=\Bra{e_r^{(n)}}C^{-1}\ket{x^{(n)}},\quad \sum_{k=1}^n \left (  C^{-1}\right )_{r,k}=\Bra{e_r^{(n)}}C^{-1}\ket{1^{(n)}}.
\end{equation}
Next, let us use Lemma \ref{vandermonde} to find that $C^{-1}=-QV_Y^{-1}V_xP^{-1}$. By a reasoning relying on the use of Lemma \ref{summation} (which is analogous to the one used in \Eref{phix}), we obtain that
\begin{equation}\label{kets}
V_xP^{-1}\ket{x^{(n)}}=
\pmatrix{
0 \cr
\vdots \cr
0 \cr
1 \cr
\sum_{k=1}^n x_k
},\quad
V_xP^{-1}\ket{1^{(n)}}=
\pmatrix{
0 \cr
\vdots \cr
0 \cr
1 \cr
}
\end{equation}
The next step is the multiplication by $V_y^{-1}$. To this end, we apply the following result concerning the inverse of a Vandermonde matrix \cite{knuth}.
\begin{lemma}\label{vandermonde-inv}
Let $V^{(n)}_t$ be the Vandermonde matrix of size $n$ in variables $t_1,\dots,t_n$, i.e. $V_{i,j}=t_i^{j-1}$, $1\leq i,j\leq n$. Then
\[\left(\left(V^{(n)}_t\right)^{-1}\right)_{i,j}=\frac{(-1)^{n-j}e_{n-j}(\{t_1,\dots,t_n\}/\{t_i\})}{\prod_{k=1,k\neq i}^{k=n}(t_i-t_k)},\]
where $e_m$ denotes the $m$th Viete sum.
\end{lemma}
\noindent We only need the last two columns of $(V_y)^{-1}$, because using \Eref{kets}, we get
\begin{eqnarray}
\Bra{e_r^{(n)}}C^{-1}\ket{x^{(n)}}=-p(y_r)\left(((V_y)^{-1})_{r,n-1}+((V_y)^{-1})_{r,n}\sum_{k=1}^n x_k\right),\\ \Bra{e_r^{(n)}}C^{-1}\ket{1^{(n)}}=-p(y_r)((V_y)^{-1})_{r,n}.
\end{eqnarray}
By Lemma \ref{vandermonde-inv}, the necessary matrix elements of $V_y^{-1}$ forms read
\[\fl ((V_y)^{-1})_{r,n-1}=-\frac{\sum_{k=1,k\neq r}^n y_k}{\prod_{k=1,k\neq r}^{k=n}(y_r-y_k)}, \quad ((V_y)^{-1})_{r,n}=\frac{1}{\prod_{k=1,k\neq r}^{k=n}(y_r-y_k)}.\]
Plugging the above results to the LHS of \Eref{identity-xi} yields
 \begin{equation*}
\sum_{k=1}^n \left (  C^{-1}\right )_{r,k}\  (   x_k -h)=\frac{p(y_r)}{\prod_{k=1,k\neq r}^{k=n}(y_r-y_k)}\left(h-y_r-\sum_{k=1}^n(x_k-y_k)\right),
\end{equation*}
which after substituting expressions for $h$ and $p(y_r)$ yields the desired result.

\section{Non-regular matrices: proof of Theorem \ref{thm:degenerate}}\label{appC}
Let us start with the simplest situation where for some $n$ we have $\left|\sigma^{(n)}\cap \sigma^{(n+1)}\right|=1$ and both spectra are non-degenerate. The other cases can be analyzed by essentially repeating and slightly adjusting the proof of the simple case.  By the interlacing property (\ref{interlace}), the scenario  described above happens if and only if there exists a single $l\in \{1,\dots,n\}$ such that either $\lambda_l^{(n)}=\lambda_l^{(n+1)}$ or $\lambda_l^{(n)}=\lambda_{l+1}^{(n+1)}$. \Eref{rlessn} applied to $k=r=l$, gives $b_{l,n+1}\xi_l^{(n)}=0$, i.e. i) $b_{l,n+1}=0$ or ii) $\xi_l^{(n)}=0$. The two cases are analyzed separately.
\begin{enumerate}
\item{$b_{l,n+1}=0$.} \Eref{rlessn} applied to $k=l$ and $r\neq l$ gives $(\lambda_l^{(n+1)}-\lambda_r^{(n)})b_{l,r}=0$, which in turn implies that $b_{l,r}$ for $r\neq l$ (recall that $\lambda_l^{(n+1)}\neq \lambda_r^{(n)}$ by the assumption that $\left|\sigma^{(n)}\cap \sigma^{(n+1)}\right|=1$). Thus, the eigenvector $\Ket{\tilde v^{(n+1)}(l)}=\Ket{v^{(n)}(l),0}$. 
\Eref{rnp1} applied for $k=l$ implies $\Braket{v^{(n)}(l)|a^{(n)}}=0$, thus $b_{l,n+1}=0$ leads to $\xi_l^{(n)}=0$.

\item{$\xi_l^{(n)}=0$ and $b_{l,n+1}\neq0$.} Here, we will show that this is in contradiction with the assumption that $\left|\sigma^{(n)}\cap \sigma^{(n+1)}\right|=1$. To this end, we combine Equations (\ref{rlessn}) for $r\neq l$ with Equations (\ref{rnp1}) to obtain the following reduced set of equations involving a Cauchy matrix
\begin{equation}
\lambda^{(n+1)}_k -\tilde h = \sum_{r=1,r\neq l}^n \frac{\left(\xi_r^{(n)}\right)^2} {\lambda^{(n+1)}_k-\lambda^{(n)}_r},  \quad\forall\ \  1\le k \le n+1,
\label {spectrum-red}
\end{equation}
where $\tilde h=\sum_{k=1,k\neq l}^{n+1}\lambda^{(n+1)}_k-\sum_{k=1,k\neq l}^{n}\lambda^{(n)}_k$. Note, that Equations (\ref{spectrum-red}) for $k\neq l$ can be viewed as a set equations for the regular spectra $\tilde\sigma^{(n-1)}:=\sigma^{(n)}/\{\lambda_{l}^{(n)}\}$ and $\tilde\sigma^{(n)}:=\sigma^{(n+1)}/\{\lambda_{l}^{(n+1)}\}$. Thus, solutions for $\left(\xi_r^{(n)}\right)^2$, $r\neq l$ read
\begin{equation}\label{xi-red}
(\xi^{(n)}_r)^2 =-\frac{\prod_{k=1,k\neq l}^{n+1}\left(\lambda^{(n)}_r-\lambda^{(n+1)}_k\right)}{\prod_{k=1,k\notin\{r,l\}}^n\left(\lambda^{(n)}_r-\lambda^{(n)}_k\right)}. 
\end{equation}
However, the above formula for $\left(\xi_r^{(n)}\right)^2$ has to be consistent with \Eref{spectrum-red} for $k=l$. By denoting $\lambda_l^{(n)}=\lambda_l^{(n+1)}:=\lambda$, this boils down to the following condition
\begin{equation}\label{lambda-cond}
\left(\lambda-\tilde h\right)=\sum_{r=1,r\neq l}^n -\frac{1} {\lambda-\lambda^{(n)}_r}\frac{\prod_{k=1,k\neq l}^{n+1}\left(\lambda^{(n)}_r-\lambda^{(n+1)}_k\right)}{\prod_{k=1,k\notin\{r,l\}}^n\left(\lambda^{(n)}_r-\lambda^{(n)}_k\right)},
\end{equation}
We view \Eref{lambda-cond} as a polynomial equation for $\lambda$. It turns out that \Eref{lambda-cond} can be simplified to the equation $\prod_{k=1,k\neq l}^{n+1}\left(\lambda-\lambda^{(n+1)}_k\right)=0$. This is in contradiction with the regularity assumption, because it implies that $\lambda=\lambda^{(n+1)}_{l+1}$.
\end{enumerate}
Summing up the above analysis of the simplest case where $\lambda_l^{(n)}=\lambda_l^{(n+1)}$ for a single $l$ and both spectra are non-degenerate, we have obtained that we necessarily have $\xi_l^{(n)}=0$ and $\Ket{\tilde v^{(n+1)}(l)}=\Ket{v^{(n)}(l),0}$. The remaining solutions for $\left(\xi_r^{(n)}\right)^2$, $r\neq l$ are given by \Eref{xi-red}. Moreover, \Eref{rlessn} applied for $k\neq l$ yields $b_{k,r}=0$ whenever $r=l$ and $k\neq l$. Thus, the expressions for the eigenvectors $\Ket{\tilde v^{(n+1)}(r)}$, $r\neq l$, of the matrix $A^{(n+1)}$ are formally identical with the expressions for the eigenvectors of a regular matrix $\tilde A^{(n)}$ obtained by removing the eigenspaces corresponding to the eigenvalue $\lambda_l^{(n)}$ in $A^{(n)}$ and $A^{(n+1)}$.

Let us next consider the situation where there is a single degeneracy block in $\sigma^{(n)}$ and $\sigma^{(n+1)}$. Note first that whenever $k\in \mc{D}^{(n+1)}(l,m)$ and $r\in \mc{D}^{(n)}(l,m)$, then \Eref{rlessn} implies that $b_{k,n+1}\xi_r^{(n)}=0$. Thus, we have two possibilities for each $k\in \mc{D}^{(n+1)}(l,m)$:
\begin{enumerate}
\item $b_{k,n+1}=0$ or
\item $b_{k,n+1}\neq 0$ and $\xi_r^{(n)}=0$ for all $r\in \mc{D}^{(n)}(l,m)$.
\end{enumerate}
As the following Lemma states, case (ii) is only relevant when $\mc{D}^{(n+1)}(l,m)=\mc{D}_{I}^{(n+1)}(l,m)$.
\begin{lemma}\label{lemma:b}
Assume that the spectrum $\sigma^{(n)}$ has degeneracy indices $\mc{D}^{(n)}(l,m)$ and denote the degenerate eigenvalue by $\lambda$. If the degeneracy indices in the spectrum $\sigma^{(n+1)}$ are of the form $\mc{D}_{II}^{(n+1)}(l,m)$, $\mc{D}_{III}^{(n+1)}(l,m)$ or $\mc{D}_{IV}^{(n+1)}(l,m)$, then for every $k\in \mc{D}^{(n+1)}(l,m)$ we have $\Braket{\tilde v^{(n+1)}(k)|e^{(n+1)}(n+1)} =0$, i.e. the coefficient $b_{k,n+1}=0$. 
\end{lemma}
\begin{proof}
({\it Ad absurdum}.) Assume that $b_{k,n+1}\neq0$ for some $k\in \mc{D}^{(n+1)}(l,m)$ and $\mc{D}^{(n+1)}(l,m)$ is of the form $\mc{D}_i^{(n+1)}(l,m)$ with $i\in\{II,III,IV\}$. Denote $\mc{K}':=\{k\in  \mc{D}^{(n+1)}(l,m):\ b_{k,n+1}\neq0\}$. Then, \Eref{rlessn} applied for any $k\in \mc{K}'$ and $r\in \mc{D}^{(n)}(l,m)$ implies that $\xi_r^{(n)}=0$ for all $r\in \mc{D}^{(n)}(l,m)$. On the other hand, for any $k\in \mc{K}'\cup\left( \{1,\dots,n+1\}/\mc{D}^{(n+1)}(l,m)\right)$,  \Eref{rlessn} gives
\[b_{k,r}=\frac{\xi_r^{(n)}}{\lambda_k^{(n+1)}-\lambda_r^{(n)}}b_{k,n+1}.\]
This in turn combined with \Eref{rnp1} for the same $k$ yields
\begin{equation}\label{set-xi-reduced}
\fl \left(\lambda_k^{(n+1)}-h\right)=\frac{\sum_{r\in\mc{D}^{(n)}(l,m)}\left(\xi_r^{(n)}\right)^2}{\lambda_k^{(n+1)}-\lambda}+\sum_{r\notin\mc{D}^{(n)}(l,m)}\frac{\left(\xi_r^{(n)}\right)^2}{\lambda_k^{(n+1)}-\lambda_r^{(n)}}.
\end{equation}
In particular, we can use the Cauchy matrix method to solve the above equations for $\sum_{r\in\mc{D}^{(n)}(l,m)}\left(\xi_r^{(n)}\right)^2$ or $\left(\xi_r^{(n)}\right)^2$ with $r\notin\mc{D}^{(n)}(l,m)$. However, each case with $D^{(n)}_i$,  $i\in \{II,III,IV\}$ has to be treated separately.
\begin{enumerate}
\item{$\mc{D}^{(n+1)}(l,m)=\mc{D}_{IV}^{(n+1)}(l,m)$.} Consider the set of Equations (\ref{set-xi-reduced}) for $k\in\{1,\dots,n+1\}/\mc{D}_{IV}^{(n+1)}(l,m)$. They are Cauchy equations for the regular spectra of a reduced dimension $\tilde\sigma^{(n-m+1)}=\{\lambda_k^{(n+1)}:\ k\in \{1,\dots,n+1\}/\mc{D}_{IV}^{(n+1)}(l,m)\}$ and $\tilde\sigma^{(n-m)}=\{\lambda\}\cup\{\lambda_k^{(n)}:\ k\in \{1,\dots,n\}/\mc{D}_4^{(n)}(l,m)\}$ with $\left(\tilde\xi_1^{(n-m)}\right)^2:=\sum_{r\in\mc{D}^{(n)}(l,m)}\left(\xi_r^{(n)}\right)^2$. In analogy to \Eref{xi-equation}, we solve \Eref{set-xi-reduced} to obtain
\begin{equation}\label{projection-norm}
\left(\tilde\xi_1^{(n-m)}\right)^2=-\frac{\prod_{k=1,k\notin\mc{D}_{IV}^{(n+1)}(l,m)}^{n+1}\left(\lambda-\lambda_k^{(n+1)}\right)}{\prod_{k=1,k\notin\mc{D}^{(n)}(l,m)}^n\left(\lambda-\lambda_k^{(n)}\right)}.
\end{equation}
As we mentioned earlier, $\mc{K'}\neq\emptyset$ implies $\left(\tilde\xi_1^{(n-m)}\right)^2=0$, thus the above expression implies that $\lambda_k^{(n+1)}=\lambda$ for some $k\notin\mc{D}_{IV}^{(n+1)}(l,m)$, which is in contradiction with the Lemma's assumptions.
\item{$\mc{D}^{(n+1)}(l,m)=\mc{D}_{II}^{(n+1)}(l,m)$ or $\mc{D}^{(n+1)}(l,m)=\mc{D}_{III}^{(n+1)}(l,m)$} Consider the set of Equations (\ref{set-xi-reduced}) for $k\in\{1,\dots,n+1\}/\mc{D}^{(n+1)}(l,m)$ with $\xi_r^{(n)}=0$ whenever $r\in\mc{D}^{(n)}(l,m)$. They are Cauchy equations for the regular spectra of a reduced dimension $\tilde\sigma^{(n-m)}=\{\lambda_k^{(n+1)}:\ k\in \{1,\dots,n+1\}/\mc{D}^{(n+1)}(l,m)\}$ and $\tilde\sigma^{(n-m-1)}=\{\lambda_k^{(n)}:\ k\in \{1,\dots,n\}/\mc{D}^{(n)}(l,m)\}$. In analogy to \Eref{xi-equation}, we solve \Eref{set-xi-reduced} to obtain
\begin{equation}\label{xis-reduced2}
\fl (\xi^{(n)}_r)^2 =-\frac{\prod_{k=1,k\notin\mc{D}^{(n+1)}(l,m)}^{n+1}\left(\lambda^{(n)}_r-\lambda^{(n+1)}_k\right)}{\prod_{k=1,k\notin\mc{D}^{(n)}(l,m),k\neq r}^n\left(\lambda^{(n)}_r-\lambda^{(n)}_k\right)},\quad r\in \{1,\dots,n\}/\mc{D}^{(n)}(l,m). 
\end{equation}
The above expressions for $(\xi^{(n)}_r)^2$ have to be compatible with \Eref{set-xi-reduced} applied to $k\in \mc{K}'$. This is in full analogy to \Eref{lambda-cond} which we treated as a polynomial equation for $\lambda$. This polynomial equation is equivalent to
\[\prod_{k=1,k\notin\mc{D}^{(n+1)}(l,m)}^{n+1}\left(\lambda-\lambda^{(n+1)}_k\right)=0.\]
This in turn implies that $\lambda_k^{(n+1)}=\lambda$ for some $k\notin\mc{D}^{(n+1)}(l,m)$, which is in contradiction with this Lemma's assumptions.
\end{enumerate}
\end{proof}
Thus, whenever $\mc{D}^{(n+1)}(l,m)\neq \mc{D}_{I}^{(n+1)}(l,m)$, we have $b_{k,n+1}=0$ for all $k\in \mc{D}^{(n+1)}(l,m)$.
Let us now focus on this particular case. By \Eref{rlessn} we get that for every $k\in \mc{D}^{(n+1)}(l,m)$ and $r\notin \mc{D}^{(n)}(l,m)$ $b_{k,r}=0$. Thus, vector $\Ket{\tilde v^{(n+1)}(k)}$ effectively belongs to the degenerate subspace of $A^{(n)}$, i.e.
\begin{equation}\label{basis}
\Ket{\tilde v^{(n+1)}(k)}=\sum_{r\in \mc{D}^{(n)}(l,m)} b_{k,r}\Ket{v^{(n)}(r),0}, \quad k\in \mc{D}^{(n+1)}(l,m).
\end{equation}
Depending on the precise form of the set $\mc{D}^{(n+1)}(l,m)$, the dimension of the degenerate subspace of $A^{(n+1)}$ can be:
\begin{enumerate}
\item equal to the dimension of the degenerate subspace of $A^{(n)}$; this happens when $\mc{D}^{(n+1)}(l,m)=\mc{D}_i^{(n+1)}(l,m)$ with $i=II,III$
\item one dimension smaller than  the degenerate subspace of $A^{(n)}$; this happens when $\mc{D}^{(n+1)}(l,m)=\mc{D}_{IV}^{(n+1)}(l,m)$.
\end{enumerate}
If $\mc{D}^{(n+1)}(l,m)=\mc{D}_i^{(n+1)}(l,m)$ with $i=II,III$, then \Eref{basis} effectively describes just a change of basis. Thus, coefficients $b_{r,k}$ with $r\in \mc{D}^{(n)}(l,m)$ and $k\in \mc{D}^{(n+1)}(l,m)$ form an orthogonal matrix. If $\mc{D}^{(n+1)}(l,m)=\mc{D}_{IV}^{(n+1)}(l,m)$, then \Eref{basis} geometrically means that the degenerate eigenspace of $A^{(n+1)}$ can be any subspace of the degenerate eigenspace of $A^{(n)}$ of codimension one. \Eref{rnp1} yields the condition that $\Braket{\tilde v^{(n+1)}(k)|a^{(n)}}=0$ for all $k\in \mc{D}^{(n+1)}(l,m)$, i.e.
\begin{equation}\label{a-ortho}
\sum_{r\in \mc{D}^{(n)}(l,m)} b_{k,r}\xi_r^{(n)}=0, \quad k\in \mc{D}^{(n+1)}(l,m).
\end{equation}
If $\mc{D}^{(n+1)}(l,m)=\mc{D}_i^{(n+1)}(l,m)$ with $i=II,III$, the dimensionality of the degenerate eigenspaces together with \Eref{a-ortho} forces $\xi_r^{(n)}=0$ for all $r\in \mc{D}^{(n)}(l,m)$. This together with \Eref{rlessn} imply that any eigenvector of $A^{(n+1)}$ corresponding to a non-degenerate eigenvalue is orthogonal to the degenerate eigenspace of $A^{(n)}$.  Thus, we have effectively reduced the dimension of the problem by $m+1$ and we can apply Theorem \ref{thm:main} for the regular reduced matrices $\tilde A^{(n-m-1)}$ and $\tilde A^{(n-m)}$ with the degenerate eigenspaces removed. In particular,
\begin{eqnarray*}
\fl b_{k,r} = \frac{\xi_r^{(n)}}{\lambda^{(n+1)}_k-\lambda^{(n)}_r}b_{k,n+1},\quad r\notin \mc{D}^{(n)}(l,m), \quad k\notin \mc{D}^{(n+1)}(l,m), \\
\fl \Ket{a^{(n)}}=\sum_{r\notin \mc{D}^{(n)}(l,m)}\xi_r^{(n)}\Ket{v^{(n)}(r),0},\quad \left(\xi_r^{(n)}\right)^2= -\frac{\prod_{k=1,k\notin\mc{D}^{(n+1)}(l,m)}^{n+1}\left(\lambda_r^{(n)}-\lambda_k^{(n+1)}\right)}{\prod_{k=1,k\notin\mc{D}^{(n)}(l,m),k\neq r}^n\left(\lambda_r^{(n)}-\lambda_k^{(n)}\right)}
\end{eqnarray*}
and $b_{k,r}=0$ for $k\notin \mc{D}^{(n+1)}(l,m)$ otherwise.

Let us next move to the case when $\mc{D}^{(n+1)}(l,m)=\mc{D}_{IV}^{(n+1)}(l,m)$. The geometric interpretation of \Eref{a-ortho} is that $\Ket{a^{(n)}}$ is orthogonal to the degenerate eigenspace of $A^{(n+1)}$ embedded in the degenerate eigenspace of $A^{(n)}$. Consider next \Eref{set-xi-reduced} from the proof of Lemma \ref{lemma:b} applied for $k\notin \mc{D}_{IV}^{(n+1)}(l,m)$. This is a set of $n-m+1$ equations involving regular spectra of a reduced dimension $\tilde\sigma^{(n-m+1)}=\{\lambda_k^{(n+1)}:\ k\in \{1,\dots,n+1\}/\mc{D}_{IV}^{(n+1)}(l,m)\}$ and $\tilde\sigma^{(n-m)}=\{\lambda\}\cup\{\lambda_k^{(n)}:\ k\in \{1,\dots,n\}/\mc{D}^{(n)}(l,m)\}$. The reduced regular matrix $\tilde A^{(n-m+1)}$ is obtained from $A^{(n+1)}$ by removing the entire degenerate eigenspace with eigenvalue $\lambda$. The reduced regular matrix $\tilde A^{(n-m)}$ is determined by the embedding of the degenerate eigenspace of $A^{(n+1)}$ into the degenerate eigenspace of $A^{(n)}$ -- the image of this embedding is removed. Next, we use the familiar method of Theorem \ref{thm:main} applied the above regular matrices. However, in contrast to the previous case the coefficients $\xi_r^{(n)}$ and $b_{k,r}$ for $r\in \mc{D}^{(n)}(l,m)$ and $k\notin \mc{D}^{(n+1)}(l,m)$ may be nonzero and are determined by the choice of the embedding of eigenspaces.

Finally, let us consider $\mc{D}^{(n+1)}(l,m)=\mc{D}_{I}^{(n+1)}(l,m)$. Firstly, note that $|\mc{D}_{I}^{(n+1)}(l,m)|=|\mc{D}^{(n)}(l,m)|+1=m+2$, thus the degenerate eigenspace of $A^{(n+1)}$ cannot be spanned only by the vectors $\Ket{v^{(n)}(r),0}$ with $r\in \mc{D}^{(n)}(l,m)$. Hence, we necessarily have $b_{k,n+1}\neq 0$ for some $k\in \mc{D}_{I}^{(n+1)}(l,m)$. By \Eref{rlessn} this implies that $\xi_r^{(n)}=0$ for all $r\in \mc{D}^{(n)}(l,m)$, i.e. $\Ket{a^{(n)}}$ is orthogonal to the degenerate eigenspace of $A^{(n)}$. This in turn implies that for any $k\notin \mc{D}_{I}^{(n+1)}(l,m)$ we have $b_{k,r}=0$ whenever $r\in \mc{D}^{(n)}(l,m)$, i.e. the non-degenerate eigenvectors of $A^{(n+1)}$ are orthogonal to the degenerate eigenspace of $A^{(n)}$. Thus, we consider the regular matrices $\tilde A^{(n-m)}$ and $\tilde A^{(n-m-1)}$ which are the original matrices truncated to the space defined as the orthogonal complement of the degenerate eigenspace of $A^{(n)}$.  They have the regular spectra $\tilde\sigma^{(n-m)}=\{\lambda\}\cup \{\lambda_k^{(n+1)}:\ k\in \{1,\dots,n+1\}/\mc{D}_{I}^{(n+1)}(l,m)\}$ and $\tilde\sigma^{(n-m-1)}=\{\lambda_k^{(n)}:\ k\in \{1,\dots,n\}/\mc{D}^{(n)}(l,m)\}$. Thus, we use Theorem \ref{thm:main} to find the vector $\Ket{a^{(n)}}$ as well as the non-degenerate eigenvectors of $A^{(n+1)}$. What is left to find are the remaining $m+2$ degenerate eigenvectors of $A^{(n+1)}$ corresponding to the eigenvalue $\lambda$. Their coefficients $b_{k,r}$ with $r\notin \mc{D}^{(n)}(l,m)$ are proportional to $b_{k,n+1}$ as stated in \Eref{rlessn}, i.e.
\[b_{k,r}=\frac{\xi_r^{(n)}}{\lambda-\lambda_r^{(n)}}b_{k,n+1},\quad r\notin \mc{D}^{(n)}(l,m), \quad k\in \mc{D}_{I}^{(n+1)}(l,m).\]
Note, that the equations
\begin{equation*}
(\lambda-h)b_{k,n+1}=\sum_{r=1, r\notin \mc{D}^{(n)}(l,m)}^n b_{k,r}\xi_r^{(n)}, \quad k\in \mc{D}_{I}^{(n+1)}(l,m).
\end{equation*}
are now automatically satisfied by the above expressions for $b_{k,r}$. The unknown coefficients $b_{k,n+1}$ and $b_{r,k}$ with $k\in \mc{D}_{I}^{(n+1)}(l,m)$ and $r\notin \mc{D}^{(n)}(l,m)$ are completely free to choose so that the resulting vectors $\left\{\Ket{\tilde v^{(n+1)}(k)}\right\}_{k\in \mc{D}_{I}^{(n+1)}(l,m)}$ form an orthonormal basis of the degenerate eigenspace.

\end{document}